\documentclass[sigconf]{acmart}
\usepackage{xspace}
\usepackage{graphicx}
\usepackage{array}
\usepackage{multirow}
\usepackage{enumitem}
\usepackage{pifont}
\usepackage{booktabs}
\usepackage[nameinlink]{cleveref}
\usepackage{amsmath}
\usepackage{algorithm}
\usepackage{algpseudocode}
\usepackage{cancel}
\usepackage{soul}

\AtBeginDocument{%
  }

\copyrightyear{2026}
\acmYear{2026}
\setcopyright{cc}
\setcctype{by}
\acmConference[CHI '26]{Proceedings of the 2026 CHI Conference on Human Factors in Computing Systems}{April 13--17, 2026}{Barcelona, Spain}
\acmBooktitle{Proceedings of the 2026 CHI Conference on Human Factors in Computing Systems (CHI '26), April 13--17, 2026, Barcelona, Spain}
\acmPrice{}
\acmDOI{10.1145/3772318.3790283}
\acmISBN{979-8-4007-2278-3/2026/04}

\begin{document}


\title{GhostUI: Unveiling Hidden Interactions in Mobile UI}
\author{Minkyu Kweon}
\orcid{0009-0000-2557-6055}
\affiliation{%
  \institution{Seoul National University}
  \city{Seoul}
  \country{Republic of Korea}
}
\email{mk@hcil.snu.ac.kr}

\author{Seokhyeon Park}
\orcid{0009-0003-1685-4027}
\affiliation{%
  \institution{Seoul National University}
  \city{Seoul}
  \country{Republic of Korea}
}
\email{shpark@hcil.snu.ac.kr}

\author{Soohyun Lee}
\orcid{0000-0002-3075-3981}
\affiliation{%
  \institution{Seoul National University}
  \city{Seoul}
  \country{Republic of Korea}
}
\email{shlee@hcil.snu.ac.kr}

\author{You Been Lee}
\orcid{0009-0000-3379-0449}
\affiliation{%
  \institution{Seoul National University}
  \city{Seoul}
  \country{Republic of Korea}
}
\email{yblee2001@snu.ac.kr}

\author{Jeongmin Rhee}
\orcid{0009-0007-3533-6603}
\affiliation{%
  \institution{Seoul National University}
  \city{Seoul}
  \country{Republic of Korea}
}
\email{jmrhee@hcil.snu.ac.kr}

\author{Jinwook Seo}
\orcid{0000-0002-7734-822X}
\authornote{Corresponding Author}
\affiliation{%
  \institution{Seoul National University}
  \city{Seoul}
  \country{Republic of Korea}
}
\email{jseo@snu.ac.kr}


\newcommand{\eg}{\textit{e.g.}\xspace}
\newcommand{\ie}{\textit{i.e.}\xspace}
\newcommand{\cf}{\textit{c.f.}\xspace}
\newcommand{\etal}{\textit{et al.}\xspace}

\definecolor{diffbg}{RGB}{242, 196, 123}
\definecolor{difftxt}{RGB}{0, 0, 255} 

\definecolor{nondiffbg}{RGB}{255, 255, 255}
\definecolor{nondifftxt}{RGB}{0,0,0}






\def\ghostuiwithpng{\raisebox{-0.2em}{\includegraphics[height=1em]{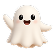}}\,\textsc{GhostUI}\xspace}
\def\ghostui{\textsc{GhostUI}\xspace}
\def\hiddeninteraction{\textit{hidden interaction}\xspace}
\def\hiddeninteractions{\textit{hidden interactions}\xspace}
\def\Hiddeninteraction{\textit{Hidden interaction}\xspace}
\def\Hiddeninteractions{\textit{Hidden interactions}\xspace}
\def\HiddenInteractions{\textit{Hidden Interactions}\xspace}
\def\HiddenInteraction{\textit{Hidden Interaction}\xspace}
\newcommand{\cmark}{\textcolor[HTML]{2ca02c}{\ding{51}}} 
\newcommand{\xmark}{\textcolor[HTML]{D62728}{\ding{55}}} 
\newcommand{\textred}[1]{\textcolor{red}{#1}}

\begin{abstract}
Modern mobile applications rely on \hiddeninteractions---gestures without visual cues like long presses and swipes---to provide functionality without cluttering interfaces. While experienced users may discover these interactions through prior use or onboarding tutorials, their implicit nature makes them difficult for most users to uncover. Similarly, mobile agents---systems designed to automate tasks on mobile user interfaces, powered by vision language models (VLMs)---struggle to detect veiled interactions or determine actions for completing tasks. To address this challenge, we present \ghostui, a new dataset designed to enable the detection of \hiddeninteractions in mobile applications. \ghostui provides before-and-after screenshots, simplified view hierarchies, gesture metadata, and task descriptions, allowing VLMs to better recognize concealed gestures and anticipate post-interaction states. Quantitative evaluations with VLMs show that models fine-tuned on \ghostui outperform baseline VLMs, particularly in predicting \hiddeninteractions and inferring post-interaction screens, underscoring \ghostui's potential as a foundation for advancing mobile task automation.
\end{abstract}



\begin{teaserfigure}
    \vspace{-1em}
    \includegraphics[width=\textwidth]{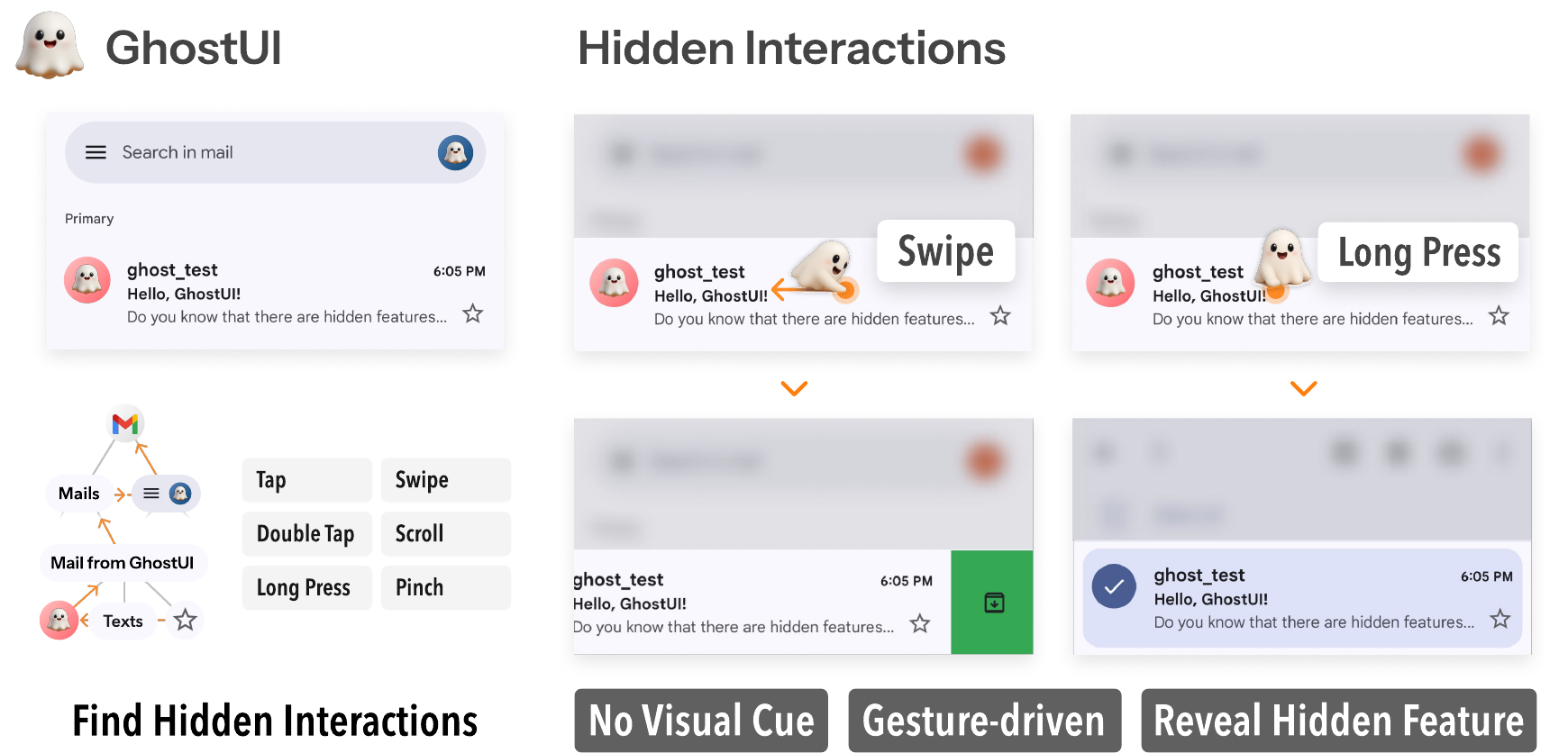}
    \caption{\ghostuiwithpng provides a dataset and framework for discovering \hiddeninteractions in mobile UIs—interactions that lack visible cues but are triggered by gestures such as swipe, long press, or double tap. The dataset systematically documents these concealed interactions through automated probing of real-world mobile applications. \Hiddeninteractions are characterized by (1) the absence of visual affordances, (2) gesture-driven activation, and (3) the revelation of previously inaccessible features.
}
    \label{fig:teaser}
    \vspace{1em}
\end{teaserfigure}


\maketitle

\section{Introduction}

Modern mobile applications support a wide variety of interaction techniques to maximize the utility of limited screen space~\cite{nilsson2009design}. While directly visible user interface (UI) elements such as buttons and menus provide clear affordances, many applications also rely on interactions without explicit visual cues---such as gestures performed on seemingly static elements. We refer to these as \hiddeninteractions: gestures like long presses or swipes that lack any visible indication that they are possible. \Hiddeninteractions serve two primary purposes: \textit{i)} providing shortcuts to frequently used functionality \textit{ii)} revealing hidden UI elements, such as menus or options, that would otherwise consume valuable screen space~\cite{cuello2013designing}. These interactions are common in modern mobile experiences. For example, in Apple's iMessage conversation list, a two-finger downward pan gesture anywhere on the message list enables quick selection of multiple messages without entering edit mode. Similarly, in Instagram's direct message screen, long pressing on an individual message reveals hidden options like deletion or relocation---features that are typically inaccessible on Android without using this gesture.

The emergence of powerful vision language models (VLMs)~\cite{zhang2024vision, ghosh2024exploring} has led to a surge in mobile agents~\cite{wu2024foundations} designed to automate UI tasks through natural language instructions, opening new possibilities for user interaction. While humans can discover and learn \hiddeninteractions through exploration, onboarding tutorials, or prior experience with similar applications, VLM-powered agents face a fundamental challenge: they can only perceive what is visually present in a screenshot. This constraint makes it difficult for such systems to perform tasks that rely on \hiddeninteractions, where the necessary actions are not visually apparent in the interface~\cite{li2024appagent}. The challenge is further complicated by the many-to-many relationship between UI elements and gestures. A single UI element may respond differently to various gestures, and the same gesture may trigger different outcomes across different elements. For instance, in the YouTube video player, a single tap reveals the playback controls, a double tap skips forward or backward, and a long press enables 2x speed playback---all on the same UI area. Conversely, the same double tap gesture produces different outcomes depending on the target element: in Instagram, double tapping a post likes the content, while double tapping a profile icon in the navigation bar switches between multiple accounts. Without clear visual affordances indicating these possibilities, VLM-based agents cannot reliably infer the correct interaction.

Existing mobile UI datasets~\cite{deka2017rico, li2020mapping, rawles2023androidinthewild, gao2024mobileviews, bai2021uibert} for training VLMs primarily focus on visually apparent interactions, such as button presses, swiping carousel, and form inputs. These datasets are typically constructed from UI screenshots and accompanying metadata, which predominantly represent elements with explicit visual cues. Although recent advances in UI understanding have made progress—with OmniParser~\cite{lu2024omniparser} improving semantic interpretation of visible components (\eg, icons, text) and Ferret-UI~\cite{you2024ferret} supporting any-resolution image analysis—they still assume that actionable elements exhibit visual affordances. As a result, both existing datasets and models for UI understanding largely ignore \hiddeninteractions---despite their critical role in modern mobile experiences. Consequently, VLM-driven agents struggle to infer and execute these gestures, significantly limiting their effectiveness in automating real-world mobile tasks that depend on these non-obvious interactions.

To address these limitations, we present \ghostui, the first comprehensive dataset specifically designed to train and evaluate VLMs on \hiddeninteractions across diverse mobile applications. In developing this dataset, we aim to answer the following research questions:

\begin{enumerate}
    \item[\textbf{RQ1}] What methodological framework can systematically uncover \hiddeninteractions across diverse mobile applications?
    \item[\textbf{RQ2}] How can we identify, categorize, and document \hiddeninteractions based on their gesture types and interaction patterns?
    \item[\textbf{RQ3}] What types of visual and structural information are critical for VLMs to accurately predict these interactions?
\end{enumerate}

To answer these questions, we first conducted a comprehensive analysis of gesture design guidelines from major mobile platforms, specifically Android and iOS. This analysis allowed us to identify a set of gesture types that are frequently used as \hiddeninteractions in a variety of applications. Based on these findings, we developed a robust data collection pipeline tailored to systematically gather diverse \hiddeninteractions from real-world mobile applications. Our pipeline executes the identified gesture types in a controlled, repeatable manner across various mobile applications, capturing comprehensive interaction data, filtering and validating data instances to retain only those that represent \hiddeninteractions, and generating detailed contextual task descriptions to reflect realistic user goals.

We organized the collected data by gesture types to enable systematic analysis of interaction patterns and gesture-element relationships across diverse contexts. This structured approach revealed common gesture patterns and characterized the nature of \hiddeninteractions in mobile interfaces, yielding critical insights about their prevalence and visual characteristics. Through comprehensive VLM experiments, we identified which interaction data types are most critical for model performance and measured improvements in the models' ability to predict and understand \hiddeninteractions. This evaluation provides quantitative evidence that \ghostui improves VLM performance on tasks involving gesture prediction and contextual UI understanding.
Lastly, we discuss potential applications enabled by \ghostui.

Our paper makes the following key contributions:
\begin{enumerate}
\item We introduce \ghostui, a dataset of 1,970 \hiddeninteraction instances collected from 81 popular mobile applications. Each entry includes paired before-and-after screenshots, simplified view hierarchies, detailed gesture metadata, and a contextual task description---structured for effective model training.

\item We establish a formal taxonomy for \hiddeninteractions in mobile applications, categorizing six distinct gesture types and their contextual usage patterns.

\item We demonstrate that VLMs fine-tuned with \ghostui significantly outperform baseline models in both gesture classification and spatial localization. Our ablation studies further reveal the relative importance of different input features for detecting \hiddeninteractions.
\end{enumerate}
\section{Related Work}

\subsection{Hidden Interactions in Mobile UIs}
Our work builds upon the concept of affordance, originally introduced in ecological psychology~\cite{gibson2014ecological}, which describes the actionable properties of objects perceived by users. Norman~\cite{norman1999affordance} adapted this concept for interface design, emphasizing the importance of perceived affordances in determining usability. Gaver~\cite{gaver1991technology} further refined the concept by distinguishing between perceptible, hidden, and false affordances—where hidden affordances represent action possibilities that exist but lack visible cues for discovery. Recent work~\cite{liao2022rediscovering, swearngin2019modeling, schoop2022predicting, wu2023never} has expanded these ideas, using data-driven and learning-based methods to detect and represent affordances in both physical and graphical user interfaces (GUIs).

Interactions with clear visual cues, such as tapping a button or an icon, typically require minimal cognitive effort due to their intuitive affordances. However, a growing number of mobile interactions rely on gestures that lack visual cues and are difficult to discover without prior experience or trial-and-error. We define these as \hiddeninteractions—interactions characterized by three properties: the absence of visual cues indicating interaction possibilities; gesture-specific activation that deviates from expected patterns for given UI elements; and functionality that is not apparent from the initial interface state. This prevalence stems from design constraints in mobile environments, where limited screen space drives designers to map multiple gestures onto single UI elements while preserving rich functionality~\cite{nilsson2009design}. While space-efficient, this approach often leaves users struggling to discover features, particularly when gestures vary across applications~\cite{munir2023usability, lee2012understanding}. As mobile applications become increasingly feature-rich and gesture-oriented, understanding and modeling these interactions becomes crucial for enhancing user experience.

Discovering \hiddeninteractions requires exhaustively testing all gestures across all interactive elements. Users cannot predict which elements respond to which gestures or what outcomes these interactions will produce without actually trying them. Accordingly, we systematically applied six gesture types to interactive elements across popular applications. We then conducted manual verification to filter out non-functional interactions and validate genuine \hiddeninteractions, combining automated exploration with human judgment to ensure accuracy. Through this semi-automated approach, we identified gesture-specific patterns revealing how \hiddeninteractions manifest in practice. Our findings provide empirical evidence for improving interaction discoverability and offer designers insights into prevalent gesture-element mappings in contemporary mobile interfaces.

\begin{table*}[!t]
\caption{Comparison of Action Spaces across Mobile Agents and Interactive Environments}
\label{tab:action_space_comparison}
\begin{tabular}{p{0.18\linewidth}>{\centering\arraybackslash}p{0.11\linewidth}>{\centering\arraybackslash}p{0.11\linewidth}>{\centering\arraybackslash}p{0.11\linewidth}>{\centering\arraybackslash}p{0.11\linewidth}>{\centering\arraybackslash}p{0.11\linewidth}>{\centering\arraybackslash}p{0.11\linewidth}}
\toprule
\textbf{System} & \textbf{Tap} & \textbf{Double Tap} & \textbf{Long Press} & \textbf{Swipe} & \textbf{Scroll} & \textbf{Pinch} \\
\midrule
\textbf{\ghostui} & \cmark & \cmark & \cmark & \cmark & \cmark & \cmark \\
\midrule
\multicolumn{7}{l}{\textit{\textbf{Mobile Agents}}} \\
ResponsibleTA~\cite{zhang2023responsible} & \cmark & \xmark & \xmark & \xmark & \xmark & \xmark \\
DroidGPT~\cite{wen2023droidbot} & \cmark & \xmark & \cmark & \xmark & \cmark & \xmark \\
AppAgent X~\cite{jiang2025appagentx} & \cmark & \xmark & \cmark & \cmark & \cmark & \xmark \\
MobileAgent E~\cite{wang2025mobile} & \cmark & \xmark & \xmark & \cmark & \cmark & \xmark \\
AutoDroid~\cite{wen2024autodroid} & \cmark & \xmark & \xmark & \cmark & \xmark & \xmark \\
VLUI~\cite{lee2024benchmarking} & \cmark & \xmark & \xmark & \cmark & \cmark & \xmark \\
MetaGUI~\cite{sun2022meta} & \cmark & \xmark & \xmark & \xmark & \cmark & \xmark \\
CogAgent~\cite{hong2024cogagent} & \cmark & \xmark & \xmark & \cmark & \xmark & \xmark \\
AutoGUI~\cite{zhang2023you} & \cmark & \xmark & \xmark & \cmark & \cmark & \xmark \\
UI-VLM~\cite{dorka2024training} & \cmark & \xmark & \xmark & \cmark & \cmark & \xmark \\
Coco-Agent~\cite{ma2024coco} & \cmark & \xmark & \xmark & \cmark & \cmark & \xmark \\
DigiRL~\cite{bai2024digirl} & \cmark & \xmark & \xmark & \cmark & \cmark & \xmark \\
SphAgent~\cite{chai2024amex} & \cmark & \xmark & \xmark & \cmark & \cmark & \xmark \\
MobileVLM~\cite{wu2024mobilevlm} & \cmark & \xmark & \xmark & \cmark & \cmark & \xmark \\
OdysseyAgent~\cite{lu2024gui} & \cmark & \xmark & \cmark & \xmark & \cmark & \xmark \\
\midrule
\multicolumn{7}{l}{\textit{\textbf{Interactive Environment}}} \\
AndroidEnv~\cite{toyama2021androidenv} & \cmark & \xmark & \cmark & \cmark & \cmark & \xmark \\
AppBuddy~\cite{shvo2021appbuddy} & \cmark & \xmark & \xmark & \xmark & \xmark & \xmark \\
Mobile-Env~\cite{zhang2023mobile} & \cmark & \xmark & \cmark & \cmark & \cmark & \xmark \\
AndroidWorld~\cite{rawles2024androidworld} & \cmark & \xmark & \cmark & \cmark & \cmark & \xmark \\
DroidTask~\cite{wen2024autodroid} & \cmark & \xmark & \xmark & \cmark & \cmark & \xmark \\
B-MoCA~\cite{lee2024benchmarking} & \cmark & \xmark & \xmark & \cmark & \cmark & \xmark \\
\bottomrule
\end{tabular}
\end{table*}

\begin{table*}[htbp]
\caption{Definitions of Gesture Types used in \ghostui.}
\label{tab:gesture_type}
\centering
\begin{tabular}{>{\raggedleft\arraybackslash}p{1.5cm}p{12.5cm}}
\toprule
\textbf{Gesture} & \textbf{Description} \\
\midrule
Tap & Single touch-and-release gesture at specific coordinates. \\
Double tap & Two consecutive taps with a brief pause between actions. \\
Long press & Extended touch at a single point, lasting for a duration. \\
Swipe & Touch movement from one position to another in a horizontal direction (left or right). \\
Scroll & Touch movement from one position to another in a vertical direction (up or down). \\
Pinch & Two-finger movement relative to a center point, either from close to far positions (zoom in) or from far to close positions (zoom out). \\
\bottomrule
\end{tabular}
\label{tab:gestures}
\end{table*}

\subsection{Mobile UI Understanding}
\paragraph{\textbf{Mobile UI Datasets}} 
Mobile UI datasets have been instrumental in advancing UI understanding. They provide large-scale training data for tasks such as UI design retrieval~\cite{deka2017rico, bunian2021vins, park2025leveraging, park2023computational}, UI element detection~\cite{xiao2024ui, gu2023mobile}, and mobile task automation \cite{song2024visiontasker, zhang2024llamatouch, wen2024autodroid, lee2023explore}.  RICO~\cite{deka2017rico} laid the foundation with over 72k screenshots and view hierarchies from Android applications, while Enrico~\cite{leiva2020enrico} added semantic annotations for design-based retrieval. Screen2Words~\cite{wang2021screen2words} introduced screen-description pairs for natural language grounding, ScreenQA~\cite{hsiao2022screenqa} provided question-answer pairs for UI reasoning, and UIBert~\cite{bai2021uibert} offered pre-training data for multimodal UI understanding. Recent datasets have targeted more complex interaction scenarios: MoTIF~\cite{burns2021mobile} studied task feasibility with sequences of both feasible and infeasible commands, AITW~\cite{rawles2023androidinthewild} captured human demonstrations for natural language-driven device control, and MobileViews~\cite{gao2024mobileviews} introduced automated collection pipelines for screen assistant tasks. ActionBert~\cite{he2021actionbert} demonstrated that user action sequences can reveal functional UI semantics beyond visual appearance. 

\paragraph{\textbf{Mobile Task Automation}}
Early research in mobile task automation focused on grounding natural language to UI actions~\cite{pasupat2018mapping, li2020mapping}, with subsequent work employing multimodal transformers~\cite{li2021vut} and interactive frameworks~\cite{li2022mug} to advance UI reasoning. The recent emergence of powerful vision language models (VLMs) like GPT-4~\cite{achiam2023gpt} and Qwen2.5-VL~\cite{bai2025qwen2} has significantly accelerated progress in this domain, providing robust screenshot understanding and GUI analysis capabilities. Building on these foundations, specialized models such as ScreenAI~\cite{baechler2024screenai}, Ferret-UI~\cite{li2024ferret}, and OmniParser~\cite{lu2024omniparser} have further enhanced UI understanding from visual, structural, and interaction perspectives. These advances have led to the development of numerous mobile agents that automate complex UI tasks through natural language instructions. Systems like AppAgentX~\cite{jiang2025appagentx}, MobileAgent-E~\cite{wang2025mobile} and CogAgent~\cite{hong2024cogagent} demonstrate diverse approaches to mobile automation.

However, our analysis of mobile agents reveals severely constrained action spaces, as shown~\Cref{tab:action_space_comparison}—none support double tap or pinch gestures, and only five systems implement long press, despite these being fundamental interactions in modern applications. This limitation stems from a critical gap in existing datasets. While datasets have evolved from basic screenshot collections to complex task demonstrations, they predominantly capture interactions with clear visual affordances and simple gestures (\eg, tap, scroll). Complex gestures like long press, double tap, and pinch remain largely undocumented, and \hiddeninteractions are entirely absent. This dataset bias directly constrains system capabilities: agents cannot learn to perform interactions they have never seen in training data. The disconnect between the rich interaction vocabulary of real-world applications and the limited action spaces in current datasets fundamentally restricts practical deployment of mobile automation systems.

To address this gap, we present \ghostui, the first dataset to systematically document \hiddeninteractions across six gesture types in 81 popular mobile applications. By expanding the action space and capturing interactions without visual cues, \ghostui enables future research toward truly comprehensive mobile UI understanding and automation. Our dataset not only reveals the prevalence of \hiddeninteractions in contemporary mobile interfaces but also provides the necessary training data for next-generation mobile agents to achieve human-like interaction capabilities.
\begin{figure*}[htbp]
    \centering
    \includegraphics[width=\textwidth]{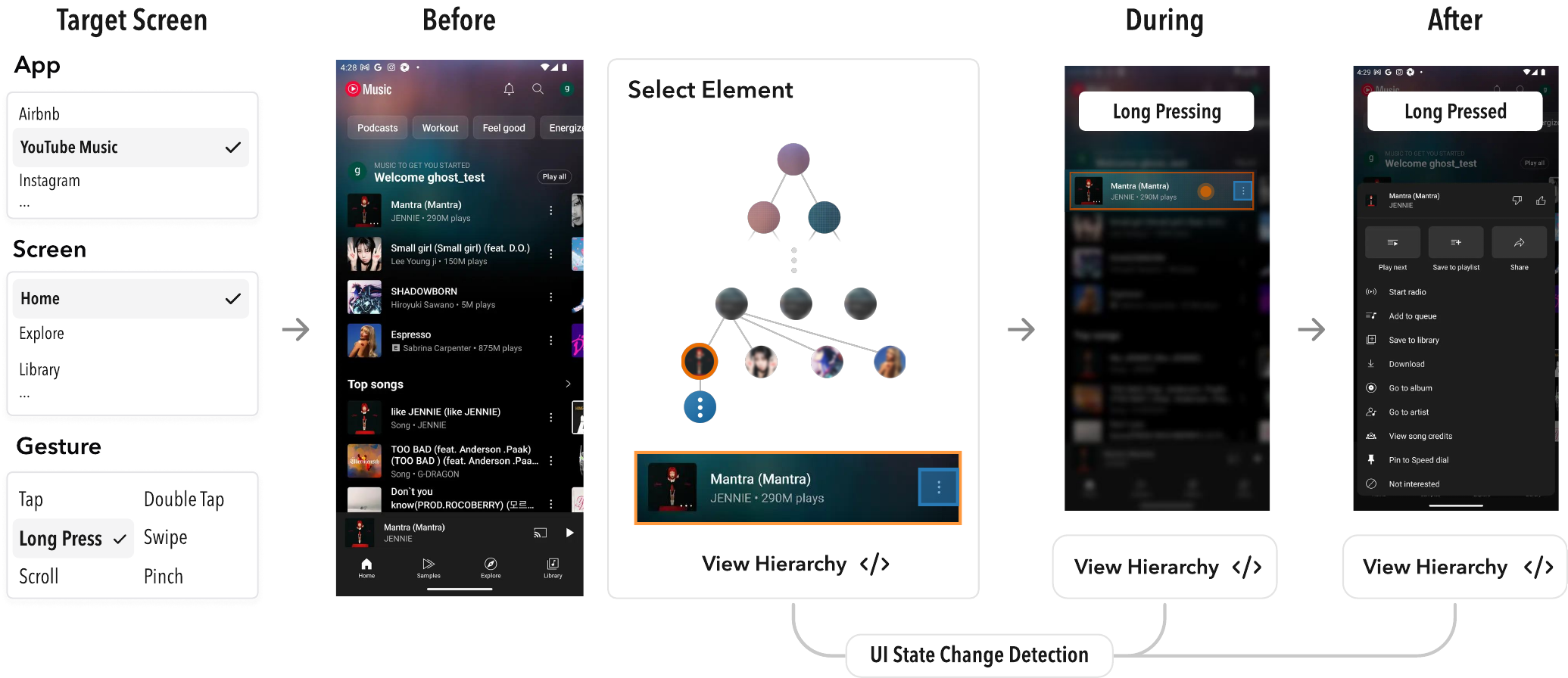}
    \caption{Overview of the UI Probing Tool Operation. \textmd{The system identifies interactive UI elements through view hierarchy parsing, tracks elements across dynamic screens using path-based identification, executes diverse touch gestures, and monitors UI states to capture ephemeral changes.}}
    \label{fig:dataset_collection}
\end{figure*}

\section{\ghostui}
\subsection{Action Space}
We first identify gesture types commonly used in mobile user interfaces by inspecting platform design guidelines from Apple Human Interface Guidelines~\cite{appleHIG} and Google Material Design~\cite{materialDesign}. We focused on six fundamental gestures that are prevalent in mobile applications and frequently trigger functionalities with no visual affordances, which highlight critical gaps in current mobile automation systems (\Cref{tab:action_space_comparison}).

While these gestures are typically applied to elements with visual cues --- tap for buttons and links, swipe on carousel indicators, scroll for scrollbars and overflowing content --- we discovered they often trigger hidden functionality in contexts without explicit affordance. For instance, as shown in~\Cref{fig:teaser}, swiping horizontally on email items in Gmail archives messages without any visual indicators suggesting this action. Similarly, long pressing the same items enables multi-selection mode despite the absence of selection checkboxes. We systematically explore these six gesture types across mobile applications to uncover such \hiddeninteractions.

\subsection{Dataset Collection Method}
To construct \ghostui, we implemented a three-phase data collection pipeline consisting of: \textit{i)} systematic interaction discovery using automated gesture testing across six gesture types, \textit{ii)} validating and filtering \hiddeninteractions through manual annotation, and \textit{iii)} task contextualization with natural language descriptions. This methodology addresses \textbf{RQ1} by establishing a scalable pipeline that effectively captures \hiddeninteractions across diverse mobile applications, yielding a task-oriented, multimodal dataset to advance intelligent mobile UI understanding.

\paragraph{App Selection Criteria}
To ensure that our dataset reflects applications users commonly encounter, we focused on popular mobile apps and selected them from the Google Play Store\footnote{\url{https://play.google.com/}}
 rankings as of March 5, 2025, starting from the top-ranked applications and proceeding in order. During this process, we excluded applications requiring subscriptions or external hardware (\eg, Netflix, Bose), streaming services whose screen content could not be reliably captured due to DRM (Digital Rights Management) protection (\eg, Pick Drama, ReelShort), and applications handling sensitive personal information, such as banking and financial services (\eg, PayPal, Cash App). Applying these criteria yielded a final set of 81 applications. Despite the exclusions, the resulting set spans diverse categories (\eg, \textit{social}, \textit{productivity}, \textit{shopping}, \textit{lifestyle}) and provides broad coverage of interaction designs commonly encountered in contemporary mobile applications.

\subsubsection{\textbf{UI Probing Tool}}
We developed an automated UI probing tool to systematically discover \hiddeninteractions across mobile applications. To facilitate reproducibility and enable crowdsourced data collection, we provide this tool as open-source on GitHub\footnote{\href{https://github.com/gh0stui/ghostui}{\texttt{https://github.com/gh0stui/ghostui}}}. The tool uses Appium\footnote{\url{https://appium.io/}} to control both an Android Emulator (Google Pixel 7 Pro) and a physical device (Samsung Galaxy A16) to capture interactions across different device environments. As shown in~\Cref{fig:dataset_collection}, this tool navigates through mobile application screens and applies every gesture to all elements of UI. To capture comprehensive UI state transitions, it records the complete before-and-after view hierarchies along with their simplified HTML-like representations, the element path, detailed gesture metadata, and corresponding screenshots for each interaction.

\paragraph{Key Screen Selection Criteria}
Our approach prioritized screens central to the app's core functionalities, rather than exhaustively covering all screens. This strategy enhanced the efficiency of our data collection process, enabling us to gather interaction data from a wider variety of applications. To implement this approach, we define key screens as primary destinations within a mobile application that are directly accessible through top-level navigation (\eg, navigation bar, tab bar). We adopted this definition from design guidelines, including Apple Human Interface Guidelines~\cite{appleHIG} and Google Material Design~\cite{materialDesign}, which recommend placing important app destinations in persistent navigation components for optimal user experience.

\paragraph{Element Detection and Tracking}
Our tool parses Android XML view hierarchies to identify interactive elements across screens. It employs a bottom-up traversal strategy, starting from leaf nodes and working upward to systematically test all elements while preventing touch point overlaps. It calculates precise bounding boxes for each element and, when determining touch points for parent nodes, carefully considers the spatial positioning of already-tested child elements to ensure non-overlapping interactions. For each element, it extracts coordinates, bounding boxes, and interactivity attributes, enabling targeted testing of elements that might support \hiddeninteractions. To handle dynamic content where UI elements change between test sessions, we implemented a path-based element tracking mechanism. This approach maps each UI element to a unique hierarchical path using class names and indices from XML view hierarchies (\eg, \texttt{FrameLayout[0]/View[0]/../Text[2]}), where each element receives a consistent index based on its position among siblings of the same type. Rather than relying on volatile properties like element size or position, this method ensures consistent identification across testing runs. By storing visited element paths in a persistent storage system classified by screen, our tool maintains consistent test coverage even when the app is restarted between testing runs or when the app's content changes. The system saves and restores test progress, allowing it to skip previously tested interactions and focus on unexplored UI elements, particularly valuable for testing dynamic applications like social media feeds where both content and spatial arrangement are frequently updated.

\paragraph{Gesture Testing and State Monitoring}
The system executes six gesture types on identified elements: tap, double tap, long press, horizontal swipes (left, right), vertical scrolls (up, down), and pinch gestures (zoom in, zoom out). For directional gestures, we capture the specific direction of each interaction---distinguishing left from right swipes, up from down scrolls, and zoom in from zoom out pinches. To ensure reproducibility, we maintain consistent interaction parameters (\eg, duration, distance) across different screens and element sizes. To detect \hiddeninteractions, we capture and compare UI states at multiple points during gesture execution. For most gestures, we record before and after states to identify UI transitions. However, for long press and pinch gestures, we additionally capture a "during" state to detect transient effects—such as tooltips that appear only while pressing a UI element or temporary zoom indicators during pinching—that would otherwise be missed.

State detection employs two complementary approaches. We primarily compare view hierarchy trees to identify structural changes in the interface, such as new elements appearing or existing ones being modified. For pinch gestures, where visual changes often occur without hierarchy modifications (\eg, image zooming), we supplement this with pixel-level screenshot comparison using a 5\% average RGB difference threshold—empirically calibrated to balance robustness against noise while maintaining sensitivity to visual changes. This dual approach ensures we capture both structural and visual \hiddeninteractions. When changes are detected, the system logs comprehensive interaction data including paired screenshots, view hierarchies, element paths, and gesture metadata with precise coordinate information—single points for taps, start-end trajectories for swipes, and multi-touch coordinates for pinches.

\paragraph{Simplified View Hierarchy Conversion}
\label{sec:simple_vh}
View hierarchies provide essential structural context for UI understanding, as demonstrated by prior work showing that models leveraging hierarchical information outperform purely visual approaches~\cite{li2020widget, bai2021uibert, li2022spotlight}. However, raw Android XML hierarchies contain excessive metadata that can overwhelm language models, these structures often exceed model context limits without improving performance. Recent work has shown that converting verbose hierarchies into simplified HTML-like representations improves model efficiency while maintaining semantic richness~\cite{wang2023enabling}. Following this approach, we developed a conversion routine that automatically transforms the captured XML view hierarchies into minimal, HTML-like structures. For each node in the XML tree, we map Android widget types (\eg, \texttt{TextView}, \texttt{RecyclerView}) to semantically simpler HTML tags (\eg, \texttt{<p>}, \texttt{<div>}) while preserving critical interaction properties (\eg, \texttt{clickable}, \texttt{bounds}, \texttt{content-desc}) as HTML attributes. This conversion process reduces noise from Android-specific attributes and creates more accessible representations for downstream analysis. As demonstrated in \Cref{sec:hidden_interaction_prediction}, this simplified view hierarchy significantly improves VLMs' spatial localization performance in \hiddeninteraction prediction tasks.

\begin{figure*}[h]
    \centering
    \includegraphics[width=\textwidth]{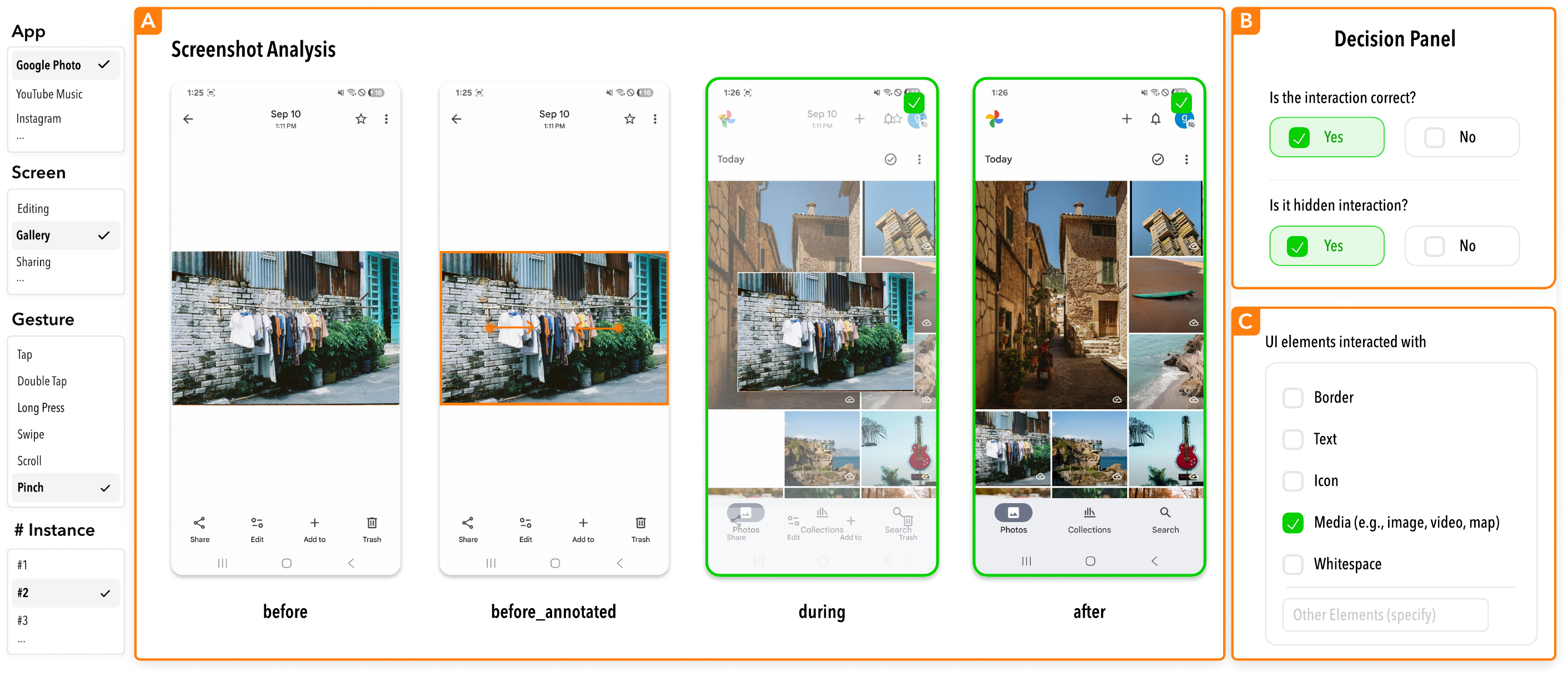}
        \caption{Overview of Validation Tool Interface for manual annotation and filtering of collected interaction data. \textmd{\textbf{(A)} Screenshot comparison panel displaying temporal UI states (before, before\_annotated with red outline indicating the target element) to facilitate visual change detection and outcome verification. \textbf{(B)} Decision panel for assessing interaction validity and determining hidden nature. \textbf{(C)} Visual element labeling panel for categorization of UI components within the target element's bounding box. In this example, only \textit{media} is selected as it represents the visual content within the target area.}}
    \label{fig:validation_tool}
\end{figure*}

\begin{figure*}[p]
    \centering
    \includegraphics[width=\textwidth, height=0.85\textheight, keepaspectratio]{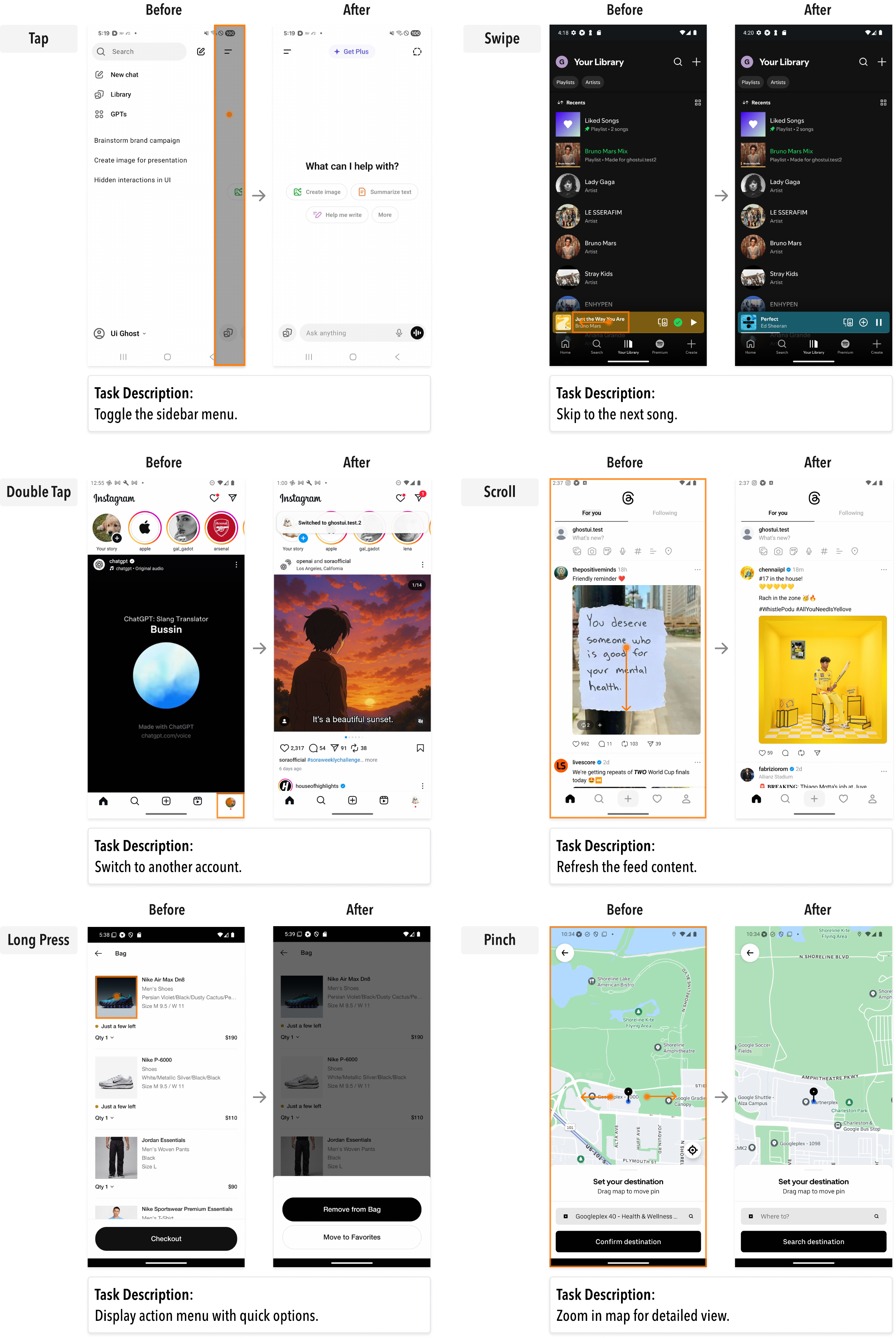}
    \caption{Examples of \HiddenInteraction across six gesture types: tap, double tap, long press, swipe, scroll, and pinch. \textmd{For each gesture type, paired before and after screenshots from real mobile apps illustrate the visual effect of interacting with a specific UI component (highlighted in orange in the before state). These examples illustrate the visual transitions that occur as a result of specific gestures applied to targeted elements.}}
    \label{fig:examples}
\end{figure*}

\begin{figure*}[t]
    \centering
    \includegraphics[width=\textwidth]{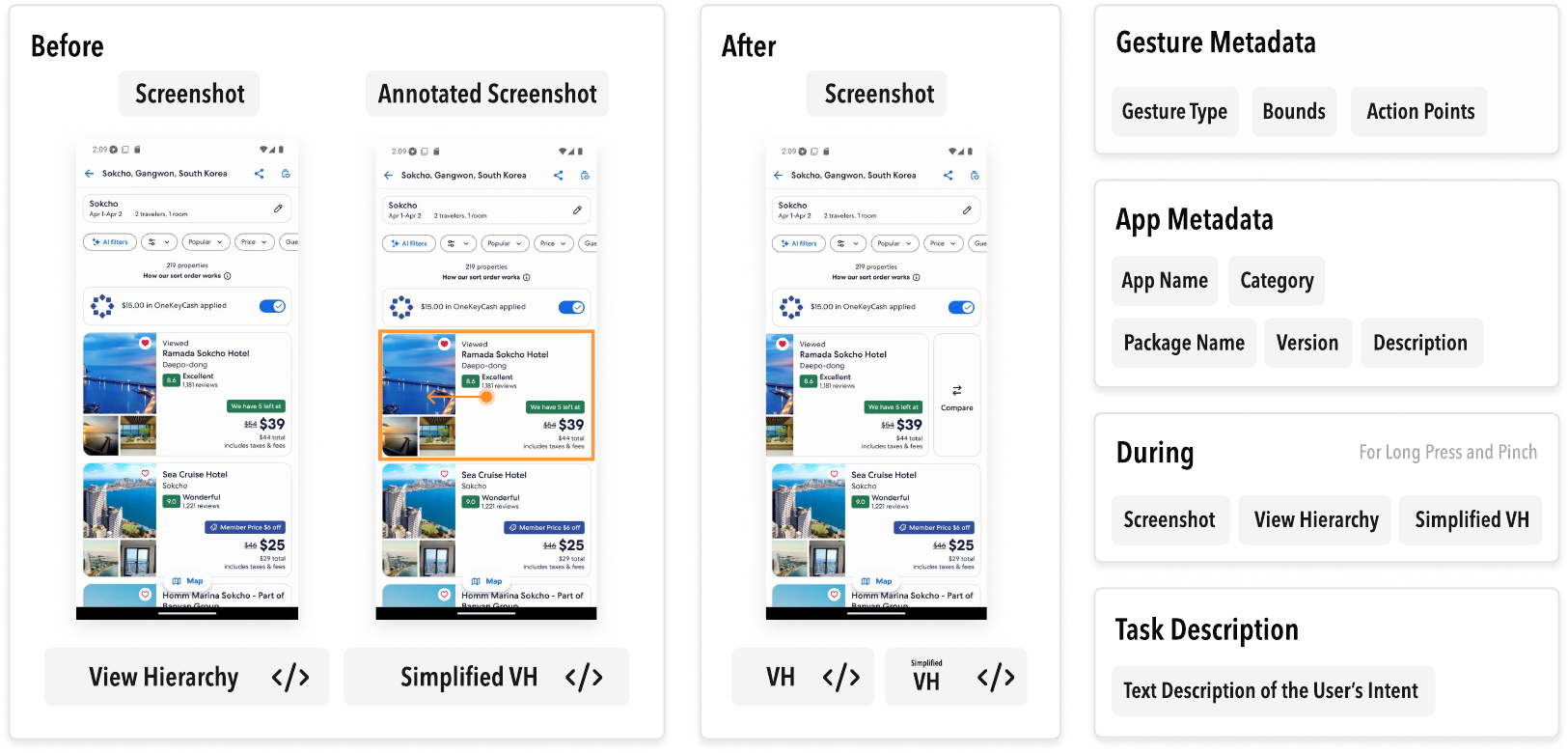}
    \caption{Structure of the \ghostui, showing the comprehensive documentation of \hiddeninteraction. \textmd{Before-and-After screenshots with view hierarchies (during state only included for long press and pinch gestures), action, app metadata, and a task description.}}
    \label{fig:datasaet_compositon}
\end{figure*}

\subsubsection{\textbf{Dataset Annotation and Validation}}
To ensure dataset quality and accurately identify true \hiddeninteractions, we developed a web-based validation tool (shown in \Cref{fig:validation_tool}) that enabled the authors to carefully review and filter all collected interaction data. We first established initial validation guidelines through consensus among five authors, then began distributed annotation. Cases that were ambiguous or difficult to classify (\eg, discussions about what constitutes a visual cue for tap-based interactions) were flagged and brought to group meetings for resolution. Through this iterative process, we refined and finalized detailed validation guidelines, which were then applied to equally divided portions of the dataset. This manual validation process followed a structured assessment framework, ensuring consistent application across the entire dataset. Along with the UI probing tool described earlier, both the validation tool and detailed annotation guidelines are made publicly available to support reproducibility.

\begin{enumerate}[left=3pt]
\item \textbf{Interaction Validity}: We verified each interaction was error-free, with no error messages or unexpected behaviors. Unexpected behaviors include unintended gesture effects, such as a double tap triggering two separate taps or a long press acting like a standard tap when applied to elements that do not support these gesture types, or swipe gestures extending beyond the target UI area. We also confirmed that the interaction produced a meaningful UI state change, where \textit{meaningful} denotes a visible change aligned with the application's intended functionality.

\item \textbf{Visual Element Labeling}: To enable systematic assessment of visual affordances, we manually labeled UI elements using five visual categories: \textit{border}, \textit{text}, \textit{icon}, \textit{media} (\eg, image, video, map), and \textit{whitespace}. This minimal set, grounded in Atomic Design~\cite{frost2016atomic} and Gestalt theories~\cite{wertheimer1938gestalt}, balances simplicity and expressiveness while capturing both content and layout-level visual cues. All elements within each gesture's bounding box were labeled using one or more of these categories, providing a structured basis for determining the visual affordance context of each interaction.

\item \textbf{Hidden Nature}: We evaluated each interaction to determine whether it lacked obvious visual cues that would typically signal interactivity. Hidden nature was assessed not only from the element itself but also from its surrounding UI context. For tap gestures, we required that the tapped element contained only \textit{whitespace} or \textit{media} without conventional affordances (\eg, borders, icons, stylized text) and that no contextual cues in the surrounding layout suggested tappability. For double tap and long press, we included cases where the gesture was applied to the same element as a standard tap but produced distinct behaviors; in such cases, the interaction was considered hidden even if visual cues were present, since the alternate behaviors could not be inferred from design alone. For swipe and scroll gestures, we confirmed the absence of contextual indicators such as cropped content, pagination dots, or scroll bars. For pinch gestures, we verified that no visual cues suggested zoom functionality, such as zoom control icons or ratio displays.
\end{enumerate}

To assess the reliability of this annotation framework, we conducted an inter-annotator agreement study. We sampled 242 interactions from the full dataset using stratified sampling to ensure proportional representation across all gesture types. All five annotators independently evaluated these samples using our established guidelines. We calculated \textit{Fleiss' Kappa}~\cite{landis1977measurement} to measure agreement across two annotation dimensions: for interaction validity and hidden nature assessment, we achieved $\kappa$=0.89 (almost perfect agreement); for visual element labeling, we achieved a mean $\kappa$=0.76 across the five element categories (substantial agreement).

Representative examples of validated \hiddeninteractions across all gesture types are provided in~\Cref{fig:examples}, illustrating the diverse visual contexts and interaction patterns captured in our dataset.

\subsubsection{\textbf{UI Task Generation}}
To contextualize \hiddeninteractions within realistic usage scenarios, we generated natural language task descriptions for each interaction in the dataset. These descriptions express the user intent and expected outcome, such as \textit{"Send a voice message in the chat"} for a long press on the voice icon in Messenger. We leveraged GPT-4o~\cite{hurst2024gpt} to analyze the before and after states of each interaction and infer the corresponding user intent. VLMs have proven effective at understanding UI semantics and generating action descriptions in recent benchmarks~\cite{baechler2024screenai, rawles2024androidworld}. The model generated descriptions based on the UI context, interaction type, and resulting state changes. We then manually reviewed these descriptions to ensure accuracy and naturalness, with the detailed prompt template provided in \Cref{appendix:task_generation}.

These task descriptions serve as multimodal supervision signals for training VLMs in affordance reasoning—learning to map high-level user intentions to concrete interaction actions. Given a visual UI state and a natural language goal, models must learn to predict both the appropriate gesture type and interaction location. This formulation enables end-to-end training for intent-to-action prediction, where VLMs learn the complex mapping between what users want to accomplish and the \hiddeninteractions required to achieve those goals. By providing this rich link between visual context, user goals, and interaction outcomes, \ghostui offers a foundation for training models to understand and execute realistic mobile interactions.

\subsection{Dataset Fields}
As illustrated in \Cref{fig:datasaet_compositon}, each data sample in \ghostui contains the following core information. The complete dataset is publicly available on HuggingFace\footnote{\href{https://huggingface.co/datasets/ghostui/ghostui}{\texttt{https://huggingface.co/datasets/ghostui/ghostui}}}.
\begin{itemize}
    \item \textbf{Screenshots:} For every interaction, we provide before and after UI screenshots, and an additional "during" screenshot for long press and pinch gestures.
    \item \textbf{View Hierarchies:} Both the complete Android XML view hierarchy and a simplified HTML-like representation, preserving key attributes such as \texttt{clickable} or \texttt{bounds}.
    \item \textbf{Gesture Metadata:} The gesture type, the corresponding bounding box, and action points.
    \item \textbf{Task Description:} A natural language statement expressing user intent, which serves as natural language supervision for training and evaluating VLMs on intent-to-interaction understanding.
    \item \textbf{App Metadata:} High-level details about the application (\eg, category, description) from Google Play Store. 
\end{itemize}

By consolidating this diverse information into each instance, \ghostui provides a robust multimodal foundation for understanding how \hiddeninteractions appear, how they alter the interface, and under what context they occur.

\begin{table*}
\caption{Distribution and Usage Patterns of \HiddenInteractions across gesture types. \textmd{The table shows frequency counts, percentages, and representative usage patterns in \ghostui.}}
\label{tab:gesture_usage_patterns_full}
\begin{tabular}{llll}
\toprule
\textbf{Gesture} & \textbf{Direction} & \textbf{Count (\%)} & \textbf{Representative Usage Patterns} \\
\midrule
Tap & - & 596 (30.3\%) & Viewing multimedia content \\
 & & & Exploring detailed information\\
 & & & Toggling navigation elements  \\
\midrule
Double Tap & - & 188 (9.5\%) & Engaging with social media content and profiles \\
 & & & Interacting with maps and location-based services \\
 & & & Searching products and services \\
\midrule
Long Press & - & 379 (19.3\%) & Opening contextual menus or hidden options \\
 & & & Selecting and managing items \\
 & & & Triggering secondary actions \\
\midrule
Swipe & Left / Right & 513 (26.0\%) & Navigating across content types or screens \\
 & & & Managing conversations and notifications \\
 & & & Exploring product categories or item lists \\
\midrule
Scroll & Up / Down & 118 (6.0\%) & Discovering new recommendations \\
 & & & Revealing extended menus or UI states \\
 & & & Browsing continuous feeds \\
\midrule
Pinch & Zoom In / Out & 176 (8.9\%) & Inspecting photos and other visual content \\
 & & & Zooming in/out on maps for spatial details \\
 & & & Adjusting perspective or layout views \\
\bottomrule
\end{tabular}
\end{table*}

\subsection{Dataset Statistics}
\label{sec:dataset_statics}
\ghostui contains 1,970 validated \hiddeninteraction instances collected from 81 popular mobile applications. Our automated UI probing tool initially gathered 8,312 interaction instances, which were then filtered through our thorough validation process to identify \hiddeninteractions. This filtering process shows that \hiddeninteractions require specific interaction patterns to discover, making them challenging for users and AI systems to identify without systematic exploration. To address \textbf{RQ2}, we systematically categorized and analyzed \hiddeninteractions based on their gesture types and interaction patterns. Our analysis examined four key dimensions: gesture type distribution across \hiddeninteractions, functional patterns and context-dependent usage across gesture types, visual element composition and co-occurrence within interactive regions, and the relationship between gesture-element combinations and interaction visibility.

\begin{figure*}[t]
    \centering
    \includegraphics[width=\textwidth]{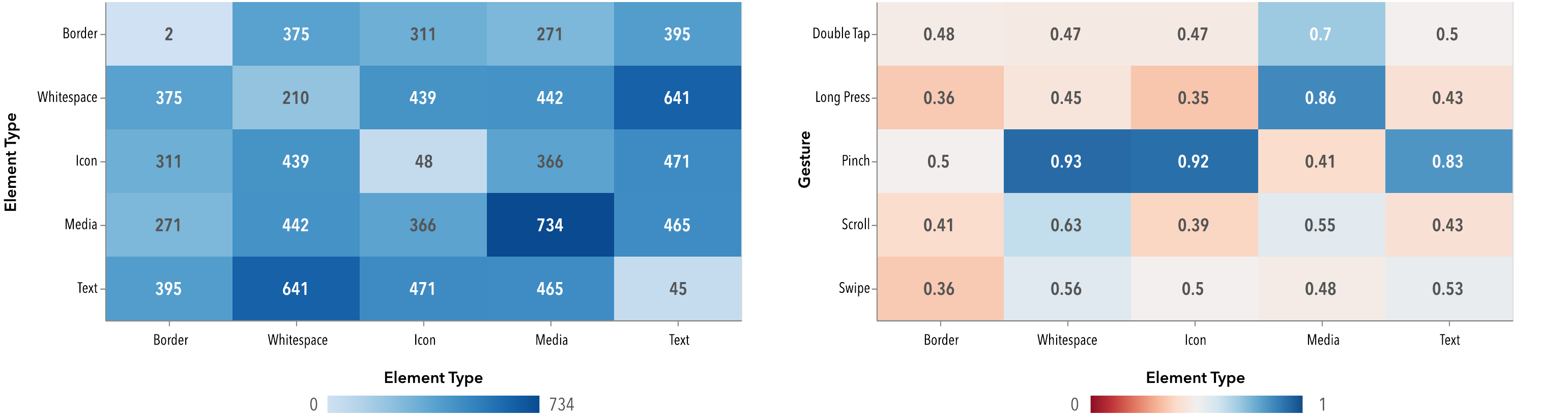}
    \caption{Element Label Co-occurrences \textbf{(left)} and Gesture-Element Distributions by element type \textbf{(right)}. \textmd{The heatmap on the left shows how often different visual labels (\eg, \textit{border}, \textit{whitespace},  \textit{icon}, \textit{media}, \textit{text}) overlap within the same bounding box, revealing frequent single-label regions (diagonal entries) and multi-label patterns (off-diagonal cells). The heatmap on the right displays each gesture type’s relative association with hidden vs. open interactions across element labels, highlighting specific label–gesture pairings more likely to indicate \hiddeninteractions.}}
    \label{fig:dataset_stat1}
\end{figure*}

\subsubsection{\textbf{Distribution of \HiddenInteractions Across Gesture Types}}
As shown in~\Cref{tab:gesture_usage_patterns_full}, \hiddeninteractions in our dataset are distributed unevenly across gesture types. Tap gestures are the most prevalent (30.3\%), representing the most fundamental input in mobile interfaces while also accounting for a substantial portion of \hiddeninteractions. Swipes (26.0\%) and long presses (19.3\%) also account for large proportions, reflecting their frequent use for navigating content and for accessing additional options such as contextual menus. Double taps (9.5\%) and pinches (8.9\%) appear less frequently overall but remain essential in specialized contexts. Scroll gestures represent the smallest portion (6.0\%), likely because scrolling is conventionally tied to continuous navigation rather than triggering hidden functionality. This distribution highlights how both fundamental and specialized gestures contribute to \hiddeninteractions across diverse mobile applications.

\subsubsection{\textbf{Gesture Usage Patterns}}
\label{sec:gesture_pattern}
The representative usage patterns in~\Cref{tab:gesture_usage_patterns_full} highlight the functional diversity of \hiddeninteractions across applications. To derive these patterns, we applied topic modeling using \textit{Latent Dirichlet Allocation} (\textit{LDA})~\cite{jelodar2019latent}, setting $k{=}3$ topics per gesture. This value was selected based on preliminary coherence tests to balance interpretability and coverage. For each task description, we extracted verb and noun phrases that capture the action–object structure of the interaction, and used them as inputs for clustering. This process yielded semantically coherent groups of interaction intents (\eg, \textit{inspect photos}, \textit{zoom maps}, \textit{manage conversations}), which we then distilled into representative usage categories for each gesture type. Our findings show both diversity and redundancy: distinct gestures often lead to similar outcomes across contexts. For example, double tap is widely used for content appreciation in social media apps, while both double tap and pinch gestures provide zooming functionality in maps and image viewers. Similarly, long press and swipe can both reveal hidden options, though their primary roles differ—long press for item selection and contextual menus, and swipe for navigation across content streams.

\subsubsection{\textbf{Element Label Composition and Co-occurrence}}
To better understand the visual context associated with \hiddeninteractions, we analyzed the semantic composition of UI elements involved in these gestures. During annotation, each target element's bounding box was labeled using five visual categories: \textit{border}, \textit{icon}, \textit{text}, \textit{media}, and \textit{whitespace}. Our probing tool applies gestures to potentially interactive elements identified from XML view hierarchy (\eg, \texttt{clickable} attributes for tap, double tap, long press). However, interactive regions often encompass multiple visual components within a single container. For instance, when only parent containers are marked as \texttt{clickable} while child elements are not, tapping different visual regions within the container may produce the same interaction outcome. The orange-highlighted component in \Cref{fig:dataset_collection} exemplifies this behavior, where tapping different areas, such as album art, title text, artist name, or surrounding whitespace, all yield the same result (excluding the blue-boxed "more" icon, which triggers a distinct action). In such cases, we assign multiple labels to regions within the bounding box that yield this shared behavior, explaining why container-based interactions frequently involve multiple visual element types.

As these labels are not mutually exclusive, we first examined how frequently different label combinations occurred together. \Cref{fig:dataset_stat1} (left) shows a heatmap representing the co-occurrence frequencies of label pairs. Diagonal cells correspond to instances where only a single label was present. Notably, \textit{media} frequently appears as the sole label, suggesting a tendency for this type to occupy isolated visual regions. In contrast, labels such as \textit{text}, \textit{icon}, and \textit{border} commonly co-occur with other types, reflecting their integration into composite UI components. This behavior is particularly common with text elements, where containers wrap both textual content and surrounding whitespace, explaining the high \textit{text}-\textit{whitespace} co-occurrence observed in our data. Surprisingly, we observed 210 instances of \textit{whitespace} being labeled alone, highlighting the prevalence of hidden gestures in visually unmarked areas.

\subsubsection{\textbf{Gesture-Element Distribution by Visibility Context}}
To further assess the affordance context of \textit{hidden} versus \textit{non-hidden} interactions, we compared the normalized gesture-element distributions across both conditions. For each gesture and element type pair, we computed the relative proportion of occurrences in hidden versus open settings. The resulting heatmap in \Cref{fig:dataset_stat1} (right) uses a diverging color scale (blue for hidden, red for open), where values closer to 1 indicate strong association with \hiddeninteractions. The visualization reveals distinct patterns in how \hiddeninteractions manifest across different visual contexts. \textit{Border} elements show the weakest association with \hiddeninteractions across all gestures, indicating that bordered elements typically provide clearer visual affordances. In contrast, elements without explicit boundaries exhibit strong \hiddeninteraction patterns: \textit{whitespace} shows high associations with hidden pinch and scroll gestures, while \textit{media} elements consistently support \hiddeninteractions for long press and double tap. Similarly, areas containing \textit{icon} and \textit{text} elements frequently support hidden pinch gestures. These findings suggest that the absence of visual boundaries correlates strongly with hidden functionality, requiring users to rely on exploratory behavior or prior knowledge to discover interactions.

\subsubsection{\textbf{Summary and Implications}}
Our analysis of \ghostui reveals that \hiddeninteractions are not edge cases but systematic patterns in mobile design. Specialized gestures (long press, double tap, pinch) constitute 37.7\% of hidden functionality yet lack both visual indicators and automation support. The container-based architecture creates ambiguous interaction targets where visually distinct regions trigger identical responses, while 210 instances of isolated \textit{whitespace} interactions demonstrate that completely unmarked areas have become legitimate targets. These empirical patterns provide a foundation for understanding how \hiddeninteractions manifest in real-world mobile interfaces and quantify the discoverability challenges that both users and automated systems face when interacting with modern mobile applications.
\section{Experiment}
In this section, we present experiments designed to evaluate the effectiveness of \ghostui in enhancing \hiddeninteraction understanding in VLMs. Our experiments address \textbf{RQ3} by \textit{i)} identifying critical information for VLMs in understanding \hiddeninteractions and \textit{ii)} validating the efficacy of our dataset through quantifiable improvements. We evaluate performance on two key tasks: \textit{Hidden Interaction Prediction}, which evaluates whether a model can identify the correct gesture type and interaction location given a UI screenshot, and \textit{UI Transition Prediction}, which measures how accurately the model can anticipate the interface changes when given a specific \hiddeninteraction. Complete prompt templates used in our experiments are provided in \Cref{appendix:prompts}.

\subsection{Experimental Setup}
\paragraph{Models}
We selected two state-of-the-art VLMs representing different model scales: Qwen2.5-VL~\cite{bai2025qwen2}, a smaller open-source model with 7B parameters, and GPT-4o~\cite{hurst2024gpt}, a larger closed-source model with significantly more parameters. For each model, we established zero-shot performance as our baseline and compared it against models fine-tuned on \ghostui. We employed Low-Rank Adaptation (LoRA)~\cite{hu2022lora} for Qwen2.5-VL and the Vision Fine-tuning API~\cite{openai2024fine-tuning} for GPT-4o.

\paragraph{Dataset Split}
To ensure robust evaluation and prevent information leakage from app-specific interaction patterns, we split the dataset such that training and test sets contain entirely different applications. Specifically, we allocated 56 apps for training and 25 apps for testing, maintaining an approximately 70:30 sample ratio while carefully balancing gesture type distributions across both splits. This app-level split ensures that models must learn generalizable patterns rather than memorizing app-specific behaviors.

\begin{table*}[!t]
\caption{\textit{Hidden Interaction Prediction} Performance (\%) across input configurations. \textmd{All-inclusive denotes the default setting with all input modalities: screenshot, view hierarchy, gesture usage pattern, and app metadata. Accuracy indicates correct gesture type classification. IoU measures spatial precision based on the overlap between predicted and ground truth bounding boxes. Relative changes from the All-inclusive setting are shown in parentheses.}}
\label{tab:interaction_prediction}
\centering
\resizebox{\textwidth}{!}{
\begin{tabular}{llcccccccccc}
\toprule
\multirow{2}{*}{\textbf{Model}} & \multirow{2}{*}{\textbf{Setting}} 
& \multicolumn{2}{c}{\footnotesize \textbf{All-inclusive}} 
& \multicolumn{2}{c}{\footnotesize \textbf{w/o View Hierarchy}} 
& \multicolumn{2}{c}{\footnotesize \textbf{w/o Gesture Pattern}} 
& \multicolumn{2}{c}{\footnotesize \textbf{w/o App Metadata}} 
& \multicolumn{2}{c}{\footnotesize \textbf{Image only}} \\
\cmidrule(lr){3-4} \cmidrule(lr){5-6} \cmidrule(lr){7-8} \cmidrule(lr){9-10} \cmidrule(lr){11-12}
& & \footnotesize \textbf{Acc.} & \footnotesize \textbf{IoU} 
  & \footnotesize \textbf{Acc.} & \footnotesize \textbf{IoU} 
  & \footnotesize \textbf{Acc.} & \footnotesize \textbf{IoU} 
  & \footnotesize \textbf{Acc.} & \footnotesize \textbf{IoU} 
  & \footnotesize \textbf{Acc.} & \footnotesize \textbf{IoU} \\
\midrule
\multirow{2}{*}{\texttt{GPT-4o}} 
& Zero-shot     
  & 51.1 & 36.0 
  & 49.6 \scalebox{0.8}{(↓1.5)} & 3.4 \scalebox{0.8}{(↓32.6)} 
  & 48.2 \scalebox{0.8}{(↓2.9)} & 35.0 \scalebox{0.8}{(↓1.0)} 
  & 50.9 \scalebox{0.8}{(↓0.2)} & 35.4 \scalebox{0.8}{(↑0.6)} 
  & 46.3 \scalebox{0.8}{(↓4.8)} & 3.5 \scalebox{0.8}{(↓32.5)} \\
& Fine-tuned   
  & 65.6 & 42.5 
  & 62.2 \scalebox{0.8}{(↓3.4)} & 4.9 \scalebox{0.8}{(↓37.6)} 
  & 57.3 \scalebox{0.8}{(↓8.3)} & 44.0 \scalebox{0.8}{(↑1.5)} 
  & 64.4 \scalebox{0.8}{(↓1.2)} & 41.8 \scalebox{0.8}{(↓0.7)} 
  & 52.8 \scalebox{0.8}{(↓12.8)} & 5.7 \scalebox{0.8}{(↓36.8)} \\
\midrule
\multirow{2}{*}{\texttt{Qwen2.5-VL}} 
& Zero-shot     
  & 33.3 & 19.5 
  & 36.8 \scalebox{0.8}{(↑3.5)} & 1.7 \scalebox{0.8}{(↓17.8)} 
  & 12.0 \scalebox{0.8}{(↓21.3)} & 23.0 \scalebox{0.8}{(↑3.5)} 
  & 34.4 \scalebox{0.8}{(↑1.1)} & 19.4 \scalebox{0.8}{(↓0.1)} 
  & 19.5 \scalebox{0.8}{(↓13.8)} & 1.9 \scalebox{0.8}{(↓17.6)} \\
& Fine-tuned   
  & 40.5 & 22.8 
  & 42.2 \scalebox{0.8}{(↑1.7)} & 6.2 \scalebox{0.8}{(↓16.6)} 
  & 36.9 \scalebox{0.8}{(↓3.6)} & 28.8 \scalebox{0.8}{(↑6.0)} 
  & 43.8 \scalebox{0.8}{(↑3.3)} & 21.2 \scalebox{0.8}{(↓1.6)} 
  & 41.7 \scalebox{0.8}{(↑1.2)} & 5.0 \scalebox{0.8}{(↓17.8)} \\
\bottomrule
\end{tabular}
}
\end{table*}

\begin{figure*}[!t]
    \centering
    \includegraphics[width=\textwidth]{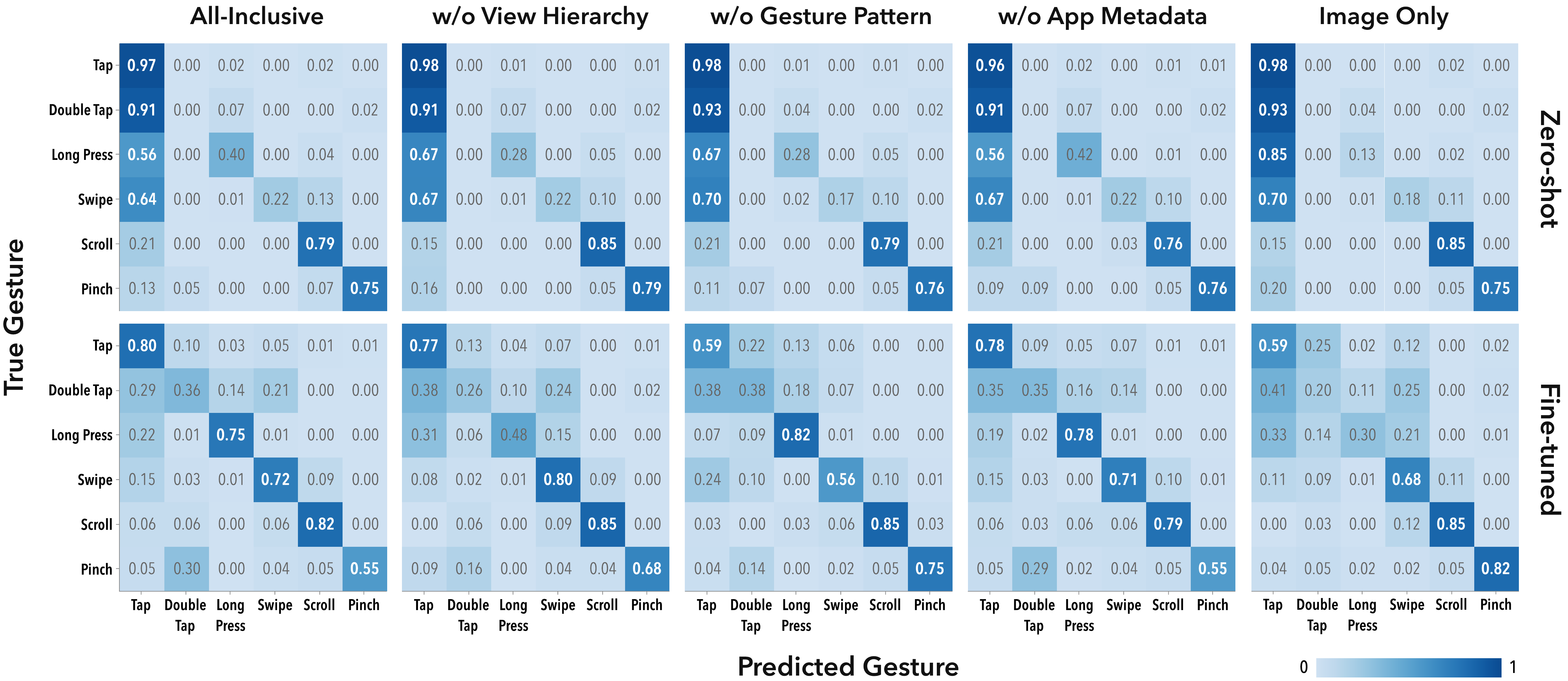}
    \caption{Confusion Matrices showing gesture type classification performance. \textmd{The matrices compare GPT-4o's performance across five input configurations in both zero-shot and fine-tuned settings. The diagonal values represent correct gesture classification probabilities, while off-diagonal values indicate confusion between gesture types. Higher values along the diagonal indicate better performance, with the fine-tuned models demonstrating significant improvements in classification accuracy compared to their zero-shot counterparts, particularly in the All-inclusive configuration (bottom-left).}}
    \label{fig:confusion_matrix}
\end{figure*}

\paragraph{Task Definition}
We consider two complementary tasks to measure how well VLMs handle \hiddeninteractions. First, \textit{Hidden Interaction Prediction} (\Cref{sec:hidden_interaction_prediction}) requires predicting which gesture (\eg, tap, long press) is needed and localizing it on the screen, given a before-interaction screenshot and a task description. Second, \textit{UI Transition Prediction}
(\Cref{sec:ui_transition_prediction}) tests whether the model can anticipate changes caused by a given \hiddeninteraction by requiring it to generate a text description of the post-gesture UI, which is then compared to a ground truth narrative.

\paragraph{Evaluation Metrics}
In \Cref{sec:hidden_interaction_prediction}, we evaluated model performance using two complementary metrics. First, classification accuracy measures whether the model correctly predicts the gesture type, including directional specifications for swipes (left, right), scrolls (up, down), and pinch gestures (zoom in, zoom out). Second, Intersection over Union (IoU) quantifies the overlap between predicted and ground truth bounding boxes to assess spatial localization precision. For \Cref{sec:ui_transition_prediction}, we measured the model's understanding of interaction outcomes by calculating cosine similarity between predicted UI descriptions and ground truth descriptions generated from after-interaction screenshots.

\begin{figure*}[t]
    \centering
    \includegraphics[width=\textwidth]{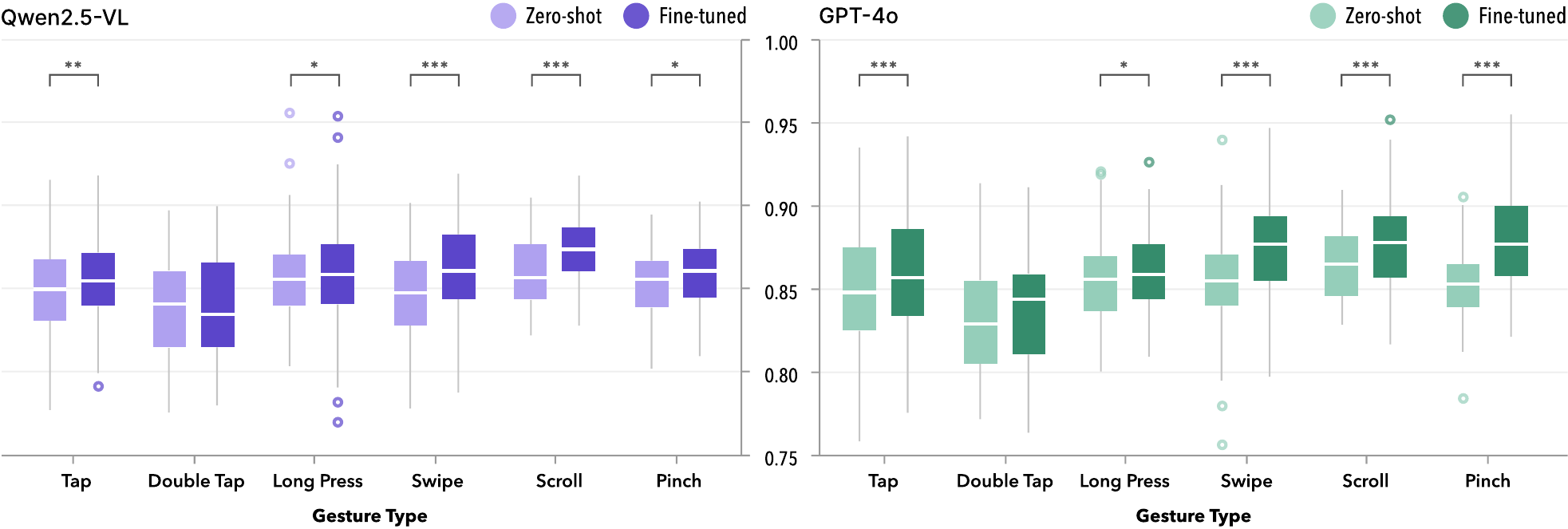}
    \caption{Cosine Similarity between predicted and ground truth after-interaction screenshot descriptions for each gesture type, comparing zero-shot and fine-tuned variants of Qwen2.5-VL (left) and GPT-4o (right). \textmd{Fine-tuning improves performance across most gesture types, with double tap being the exception. Boxes represent interquartile ranges, with medians shown as horizontal bars. Statistical significance between zero-shot and fine-tuned variants is indicated by asterisks (*$p<0.05$, **$p<0.01$, ***$p<0.001$).}}
    \label{fig:cosine_similarity}
\end{figure*}

\subsection{Hidden Interactions Prediction}
\label{sec:hidden_interaction_prediction}
To identify which input components contribute most significantly to VLM performance for \hiddeninteraction prediction, we conducted a comprehensive ablation study followed by fine-tuning experiments.

\subsubsection{\textbf{Experimental Design}}
Models were provided with a task description and before-interaction screenshot, then asked to predict the appropriate gesture type (\eg, scroll down, pinch zoom in) and the precise location where the gesture should be applied to complete the task successfully. We designed our experiments around 5 distinct input configurations to isolate the contribution of each information type. Our complete configuration (\textit{All-inclusive}) included before screenshot, simplified view hierarchy, gesture usage patterns described in~\Cref{sec:gesture_pattern}, and app metadata. We then removed individual resources to create four additional configurations: one without the simplified view hierarchy, another without gesture usage patterns, a third without app metadata, and finally a minimal configuration with only before screenshot (\textit{Image only}).

\subsubsection{\textbf{Results}}
As shown in~\Cref{tab:interaction_prediction}, our experiments revealed several key insights into \hiddeninteraction prediction. In zero-shot settings, GPT-4o demonstrated superior performance compared to Qwen2.5-VL across all input configurations, achieving 51.1\% accuracy and 36.0\% IoU with the complete information sources (\textit{All-inclusive}), while Qwen2.5-VL reached 33.3\% accuracy and 19.5\% IoU. The removal of simplified view hierarchy information resulted in a substantial decrease in spatial prediction accuracy, with IoU dropping by 32.6\% for GPT-4o and 17.8\% for Qwen2.5-VL, while gesture classification accuracy remained relatively stable. This indicates the critical importance of structural UI information for precisely localizing interaction targets. Notably, the absence of gesture pattern information had particularly severe consequences for Qwen2.5-VL's performance, causing accuracy to plummet from 33.3\% to 12.0\%. This suggests that smaller models rely more heavily on explicit gesture contextual information to make accurate predictions. 

After fine-tuning on \ghostui, both models exhibited notable performance improvements. In the \textit{All-inclusive} configuration, GPT-4o achieved 65.6\% gesture classification accuracy and 42.5\% IoU, while Qwen2.5-VL reached 40.5\% and 22.8\% respectively. The performance gains over zero-shot baselines (14.5\% accuracy for GPT-4o, 7.2\% for Qwen2.5-VL) demonstrate that \ghostui provides transferable knowledge for predicting \hiddeninteractions in unseen applications. Removing gesture usage patterns caused an 8.3\% accuracy drop for fine-tuned GPT-4o (65.6\% to 57.3\%), indicating that gesture conventions serve as valuable prior knowledge. Removing the simplified view hierarchy continued to severely impact spatial localization across both models, while removing app metadata had minimal effect, suggesting that high-level application descriptions provide little actionable information for \hiddeninteraction prediction.

To better understand how well the model distinguishes between individual gesture types, we analyzed gesture-wise confusion matrices for GPT-4o across five input configurations in~\Cref{fig:confusion_matrix}. We focused on GPT-4o for this in-depth analysis because it demonstrated consistently superior performance compared to Qwen2.5-VL on \hiddeninteraction prediction, allowing us to examine subtle patterns in gesture type classification without being confounded by overall model performance issues. 

In zero-shot settings, the model exhibited a severe bias toward predicting tap gestures. Across all configurations, double tap instances were misclassified 
as tap over 91\% of the time, and long press instances showed 56--85\% tap misclassification rates depending on input configuration. Scroll and pinch gestures were relatively better preserved even in zero-shot settings, achieving 76--85\% and 75--79\% accuracy respectively, likely because their distinct interaction patterns provide clearer differentiation signals.

Fine-tuning alleviated this tendency, leading to more balanced and accurate predictions across gesture types. In the \textit{All-inclusive} configuration, double tap recognition improved from 0\% to 36\%, long press accuracy increased from 40\% to 75\%, and swipe recognition improved from 22\% to 72\%. The visualization in ~\Cref{fig:confusion_matrix} reveals a progressive improvement in diagonal pattern clarity as we move from zero-shot to fine-tuned models and from limited to more comprehensive input configurations. This strengthening of the diagonal elements visually confirms the model's increasing ability to correctly distinguish between different gesture types. The fine-tuned models show markedly clearer diagonal patterns compared to their zero-shot counterparts, demonstrating that task-specific training on \ghostui significantly improves the model's ability to differentiate between gesture types regardless of the input configuration used.

However, certain challenges persisted even after fine-tuning. Double tap remained the most difficult gesture to classify (36\% accuracy), with substantial confusion toward tap (29\%) and swipe (21\%). Pinch gestures showed notable confusion with double tap (30\% misclassification), likely because both gestures often target similar visual contexts such as images and maps. The confusion matrices also reveal configuration-specific patterns: without view hierarchy, long press accuracy dropped from 75\% to 48\% with increased tap confusion; without gesture patterns, swipe recognition degraded from 72\% to 56\%. The image-only configuration showed the most severe degradation for long press (30\% accuracy), confirming that this gesture type is particularly dependent on structural information.

\subsection{UI Transition Prediction}
\label{sec:ui_transition_prediction}
While \Cref{sec:hidden_interaction_prediction} focuses on identifying appropriate gestures and their locations, \textit{UI Transition Prediction} evaluates whether models can accurately anticipate the interface changes that result from performing these interactions. This task provides a complementary measure of \hiddeninteraction understanding, as it requires models to comprehend not only where and how to interact, but also the consequences of those interactions. Understanding gesture consequences is a key component of comprehensive interaction reasoning, as effective planning depends on anticipating post-gesture outcomes—a critical capability for real-world mobile agents.

\subsubsection{\textbf{Experimental Design}}
In this experiment, models were provided with a before-interaction screenshot and action details (gesture type and bounding box location), then tasked with predicting the visual changes that would result from performing the specified gesture. We evaluated predictions by comparing them against descriptions of the actual after-interaction screenshots. Following prior work that demonstrates GPT-4's ability to generate accurate and fine-grained UI descriptions comparable to human annotations~\cite{wu2023gpt4vis, wang2023enabling, baechler2024screenai, park2025leveraging}, we used GPT-4o to generate ground truth descriptions from the after-interaction screenshots. To reduce sensitivity to prompt variations and description detail differences, we measured performance by computing the cosine similarity between text embeddings of the predicted and ground truth descriptions using Google's \texttt{gemini-embedding-001} model with \textit{semantic similarity} mode. This metric focuses on semantic alignment rather than exact lexical matching, where higher values indicate better understanding of gesture effects.

\subsubsection{\textbf{Results}}
As shown in~\Cref{fig:cosine_similarity}, fine-tuning on \ghostui improved both models' ability to predict interface changes resulting from \hiddeninteractions across most gesture types. We conducted a Wilcoxon Signed-rank test~\cite{wilcoxon1945individual} to assess statistical significance (*$p<0.05$, **$p<0.01$, ***$p<0.001$). Both models showed significant improvements for five of six gesture types: swipe and scroll ($p < 0.001$), tap ($p < 0.001$ for GPT-4o, $p < 0.01$ for Qwen2.5-VL), pinch ($p < 0.001$ for GPT-4o, $p < 0.05$ for Qwen2.5-VL), and long press ($p < 0.05$).

Double tap was the only gesture type showing no significant improvement ($p > 0.05$). Analysis of model outputs revealed that zero-shot models exhibited a strong bias toward predicting zoom-related outcomes, despite zoom being rare in our test set's ground truth. Additionally, double tap showed high outcome diversity, with context-dependent results varying from content engagement to text selection. These factors made it difficult for fine-tuning to overcome pre-existing biases. Overall, these findings indicate that fine-tuning on \ghostui improves gesture-induced UI transition prediction, though gestures with high outcome diversity and strong pre-training biases remain challenging.
\begin{figure*}
    \centering
    \includegraphics[width=\textwidth]{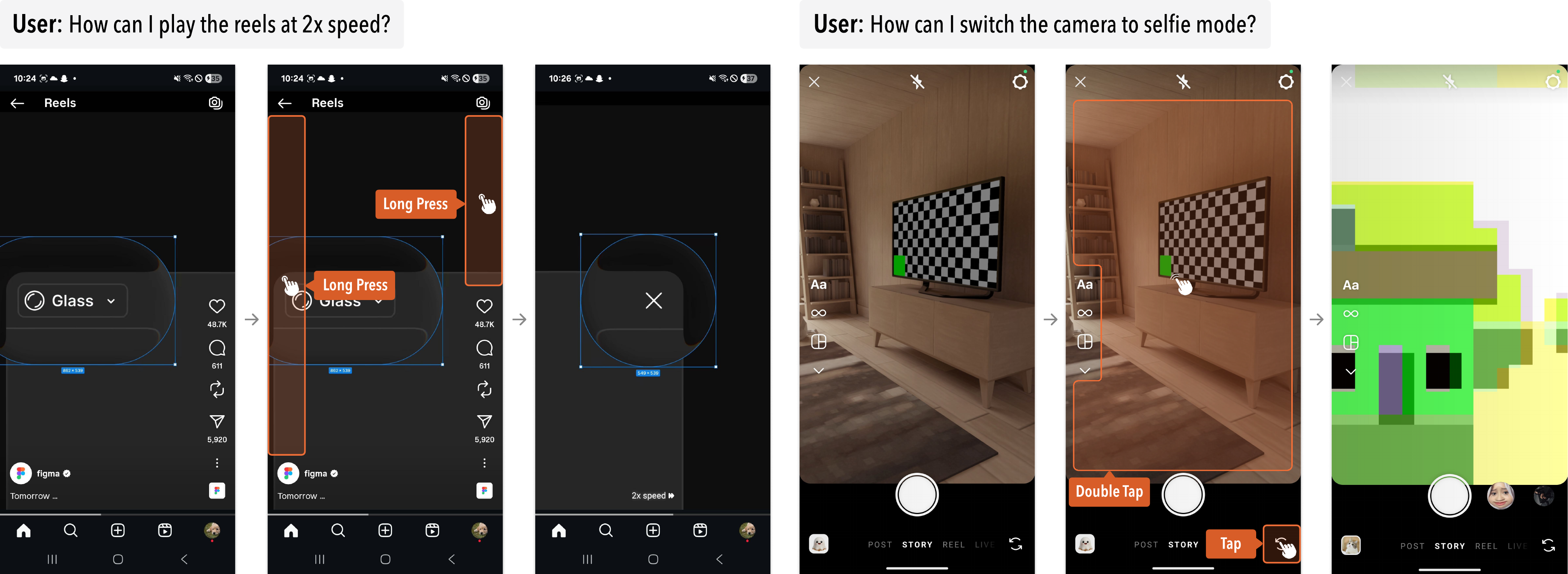}
    \caption{Illustrations of Interactive Guidance for discovering \hiddeninteractions based on user queries. \textmd{\textbf{(Left)} Long press gesture on specific regions of the video area to activate 2x playback speed. \textbf{(Right)} Double tap gesture on camera screen to switch to selfie mode, with alternative tap gesture shown for comparison.}}
    \label{fig:potential_1}
\end{figure*}

\begin{figure*}
    \centering
    \includegraphics[width=\textwidth]{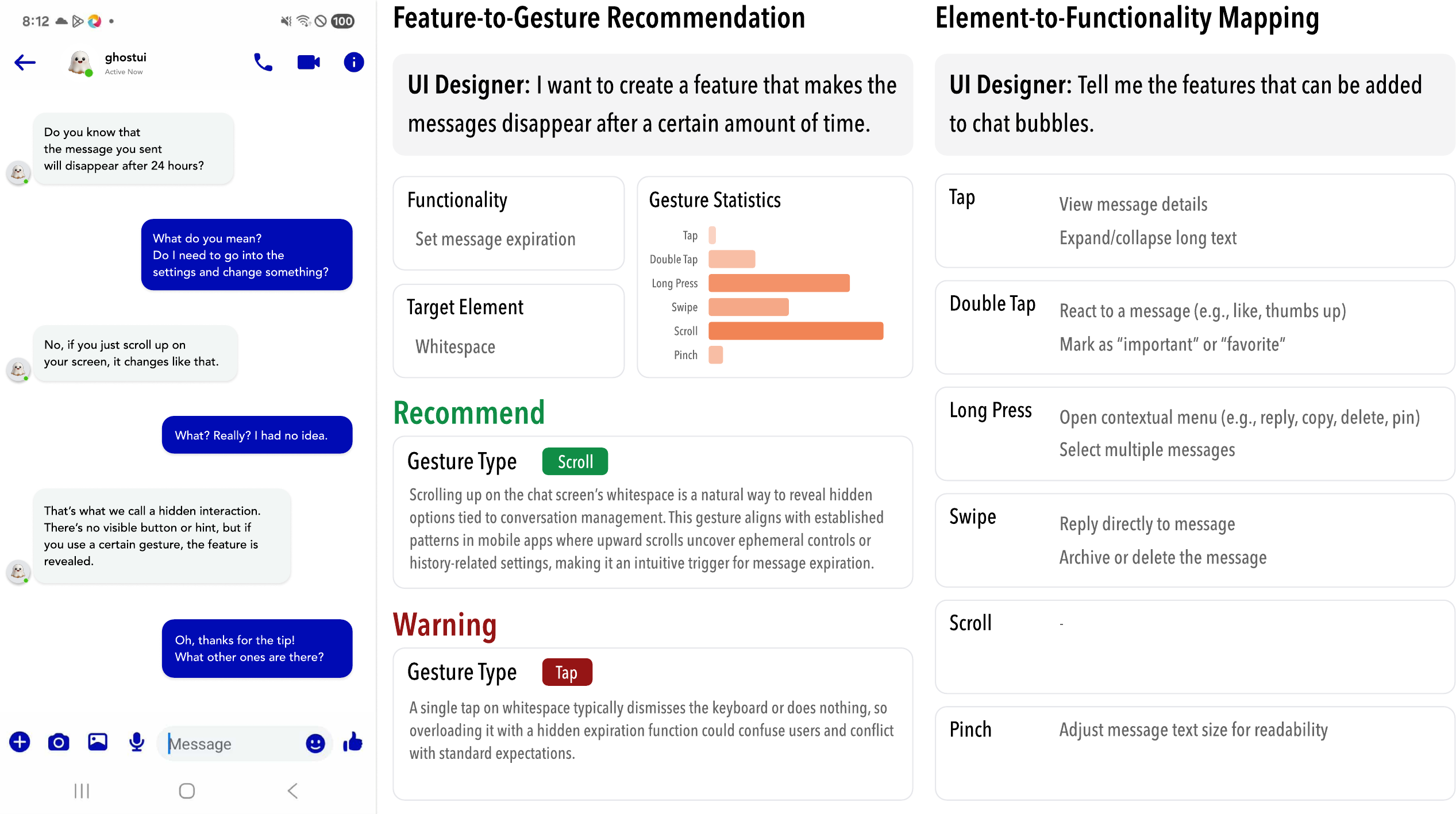}
    \caption{Illustrations of Design Recommendations for \HiddenInteractions in messaging apps. \textmd{\textbf{(Left: Feature-to-Gesture Recommendation)} The system analyzes a disappearing message feature and provides gesture suggestions based on empirical patterns, highlighting recommended options as well as potential pitfalls. \textbf{(Right: Element-to-Functionality Mapping)} The system presents mappings between gestures and their empirically observed functionalities in chat bubble interactions, such as viewing details, reactions, contextual menus, and quick replies or archiving.}}
    \label{fig:potential_2}
\end{figure*}

\section{Potential Applications}
Based on our demonstration that models trained on \ghostui can effectively predict gesture type, target region, and ensuing UI state changes, this section describes mock-up applications enabled by \ghostui, highlighting its potential across user guidance and interaction design.

\subsection{Interactive Guidance System for Hidden Interactions}
Users often struggle to discover hidden features in mobile apps, missing functionality that could enhance their experience. An interactive guidance system powered by \ghostui could reveal these interactions through contextual assistance. As shown in Figure~\ref{fig:potential_1}, when a user asks \textit{"How can I play the reels at 2x speed?"}, the system identifies that long pressing on the left or right edges of the video immediately plays it at 2x speed while held, displaying translucent overlays to guide the interaction. Similarly, for camera mode switching (right), the system reveals that double tapping anywhere on the camera screen switches to selfie mode—a hidden alternative to the visible flip icon that users might not be aware of. This approach preserves clean interfaces while ensuring users can discover hidden functionality precisely when they need it, following progressive disclosure principles where hints appear contextually and fade after successful execution.

\subsection{Interaction Design Recommendations}
Designers currently lack data-driven methods to evaluate whether their \hiddeninteraction implementations align with user expectations and platform conventions. \ghostui enables automated design conformance analysis by comparing implementations against established patterns from popular apps. As illustrated in Figure~\ref{fig:potential_2}, when a designer wants to implement disappearing messages (left), the system analyzes the target element (\textit{whitespace}) and recommends a scroll-up gesture based on empirical patterns where upward scrolls commonly reveal ephemeral controls. It warns against using tap on \textit{whitespace}, as this gesture typically dismisses keyboards rather than triggering hidden features, potentially confusing users. For adding features to chat bubbles (right), the system provides a comprehensive mapping of gestures to their appropriate functionalities: tap for viewing message details, double tap for reactions, long press for contextual menus, and swipe for quick reply or archive actions. This empirical feedback helps designers make informed decisions that balance innovation with familiarity, ensuring \hiddeninteractions remain discoverable through consistency with established patterns.
\section{Discussions}
\ghostui provides a foundation for understanding \hiddeninteractions in mobile UIs, while also opening numerous avenues for future research. In this section, we discuss the implications of our findings and outline promising directions to extend our work.

\subsection{Advanced Mobile Agents}
In \ghostui, 23.7\% of all validated interactions are classified as \hiddeninteractions, underscoring the substantial portion of mobile functionality that current agents struggle to address. Among these \hiddeninteractions, 37.7\% rely on gesture types rarely or never supported by existing mobile agent frameworks, such as double tap, long press, and pinch. The limitations are twofold: current agents face difficulty detecting \hiddeninteractions in general, and a considerable fraction remain entirely inaccessible due to unsupported gesture types.

Our experiments also highlight the importance of multimodal input. While recent GUI agents have increasingly adopted vision-centric approaches that rely solely on screenshots~\cite{xu2024aguvis, gou2024navigating}, our findings reveal a fundamental limitation of such methods for \hiddeninteractions. Removing simplified view hierarchy information caused IoU to drop dramatically, while classification accuracy remained relatively stable. This disparity stems from a mismatch between visual and interactive boundaries: VLMs tend to localize visually salient objects such as text or icons, whereas actual touch targets in mobile interfaces often encompass entire containers including surrounding whitespace and padding. For conventional UI elements like buttons, visual boundaries approximate interactive boundaries, enabling visual grounding approaches such as SeeClick~\cite{cheng2024seeclick} and OmniParser~\cite{lu2024omniparser} to succeed. However, these methods are trained and evaluated exclusively on visible elements. \Hiddeninteractions lack visual affordances by definition, making their interactive boundaries invisible to vision-only methods. Accurate localization of such interactions requires structural information that captures interactive boundaries independent of visual appearance, whether obtained directly from system-level view hierarchies or learned from specialized datasets like \ghostui that explicitly document these invisible targets.

Taken together, these results point to direct design implications for next-generation mobile agents. Expanding the action space ensures that agents can encompass the full range of gestures observed in practice, while multimodal input equips them to resolve ambiguities and localize targets more effectively. For practical deployment, \ghostui-trained models can augment existing agent architectures as specialized perception modules. The model should be invoked at three key decision points during agent execution: after initial screen analysis when no visible elements match the task intent, when action confidence scores fall below predefined thresholds, or when task descriptions explicitly reference gesture-based interactions (\eg, \textit{"long press to select"}). Integration naturally occurs within the perception layer of current frameworks---positioned between screen parsing and action planning stages---where the model can be accessed as a tool through standardized protocols such as Model Context Protocol (MCP). These advances enable agents to cover workflows that are currently inaccessible, reduce multi-step reasoning chains into single-step predictions, and improve overall task success rates. \ghostui thus provides both empirical evidence of current limitations and a foundation for systematically developing and benchmarking more capable mobile agents.

\subsection{Usability and Interaction Design Implications}
Our analysis of gesture usage patterns reveals both diversity and redundancy in how gestures contribute to application functionality. We identified representative patterns across apps, yet many gestures also enable app-specific functions that are not fully captured by broader cross-app clusters. This highlights the dual character of gesture usage: while certain patterns recur across applications, others serve highly specialized roles, reflecting the variability of design practices in modern mobile ecosystems. The element-level labeling analysis further clarifies how different visual contexts relate to gesture activation. By examining label composition and co-occurrence, as well as gesture–element distributions by visibility context, we identified consistent associations between gestures and their target elements. For example, double taps are frequently tied to image or video regions and borders to visible and thus non-hidden interactions. These associations provide a systematic view of the affordance structures underlying mobile interfaces, offering empirical evidence of where discoverability breaks down.  

These patterns raise an important question for end users: do \hiddeninteractions serve as convenient shortcuts, or do they represent the sole pathway to functionality? To explore this, we extracted recurring intents from task descriptions, selected 200 interactions through stratified sampling across intent categories, and manually examined alternative pathways in a sample of interactions. Our analysis revealed significant variation across applications. In many cases, \hiddeninteractions provided efficient shortcuts to functionality also accessible through visible UI elements---for instance, in Etsy, long pressing an item reveals options for reporting, adding to collections, and sharing, all of which are also available through buttons on the item's detail page. However, certain features remained exclusively hidden; emoji reactions in chat messages consistently lacked visible triggers across applications, unlike reactions in feeds which typically display like buttons. Most concerning were cases where alternatives themselves were \hiddeninteractions---Band allows message replies through either long press or swipe, both lacking visual cues. Such redundant hidden pathways provide no benefit for discoverability. When \hiddeninteractions lack visible alternatives, users must rely on prior experience, external documentation, or trial-and-error exploration to access features. This challenge is amplified for users with motor impairments or those relying on assistive technologies that do not readily support complex touch gestures. The prevalence of exclusively hidden functionality---and especially cases where even the alternatives remain hidden---highlights a systematic discoverability gap in contemporary mobile applications.

For designers and developers, \ghostui offers a data-driven foundation for determining when \hiddeninteractions align with conventional usage patterns and when they diverge. By grounding design decisions in empirical patterns from widely-used applications, they can audit their applications for discoverability gaps, anticipate user expectations, and identify cases where core functionality should be made available through more accessible alternatives. In this way, \ghostui supports both improved user experience and principled design innovation, ensuring that the trade-off between interface minimalism and feature accessibility is approached in an evidence-based manner.

\subsection{Improving Dataset Coverage and Scalability}
Our collection strategy focused on key screens---primary destinations accessible through top-level navigation—to efficiently cover many applications. However, this approach inherently limited exploration of nested interfaces where additional \hiddeninteractions may exist. The cold-start problem further constrained our coverage: in freshly installed apps, generating sufficient usage history required considerable time, delaying access to features that only emerge after regular use, such as chat history interactions or personalized recommendations triggered after viewing several videos. Moreover, manual annotation of \hiddeninteractions remains time-consuming, as each collected instance requires careful human verification to ensure quality.

Looking forward, autonomous agents could address the cold-start problem by navigating through multiple screens and building up usage history. If agents could also handle tedious tasks such as login and sign-up, we could focus our time on covering more screens and discovering \hiddeninteractions in deeper parts of applications. We experimented with mobile automation frameworks that enable LLMs to interact with mobile applications, though current performance remains insufficient for reliable autonomous exploration. For the annotation bottleneck, semi-automatic visual element labeling (\eg, VLM-assisted pre-labeling with human confirmation) and lightweight crowdsourced validation could reduce verification overhead while maintaining reliability. As these technologies mature, combining agent-driven exploration with VLM-assisted annotation could transform \ghostui's collection from a semi-automated process with limited scope to an automated, comprehensive system that continuously discovers \hiddeninteractions across the full depth of mobile applications.

\subsection{Limitations and future work}
\paragraph{Platform and Device Coverage.}
The current version of \ghostui focuses exclusively on Android mobile applications. While this scope enabled systematic data collection, it does not capture the broader diversity of platforms and interaction paradigms in contemporary computing. Future extensions should consider other operating systems, particularly iOS with its distinct gesture conventions, and additional device form factors. Wearable devices with constrained displays present particularly compelling targets for investigation, as their limited screen real estate likely necessitates even greater reliance on \hiddeninteractions for accessing functionality.

\paragraph{Interaction Type Coverage.}
Beyond the six gestures currently documented, numerous complex interaction types remain unexplored. Drag-and-drop operations and multi-finger gestures represent additional interaction patterns that our dataset does not address. Furthermore, sensor-based interactions—such as shaking, tilting, or rotating the device—constitute another category of hidden functionality that merits future investigation. Expanding the action space to encompass these richer interaction types would enable more comprehensive automation capabilities and better reflect the full spectrum of user behaviors in real-world applications.

\paragraph{Human Factors Research.}
While our dataset quantifies the prevalence and distribution of \hiddeninteractions, it does not address fundamental questions about human cognition and behavior. Understanding the discoverability and learnability of \hiddeninteractions requires dedicated user studies that examine how people encounter, internalize, and recall these gestures over time. Human-centered research investigating these aspects could inform design guidelines that balance the space-efficiency benefits of \hiddeninteractions with their inherent accessibility challenges, complementing our technical contributions with insights into user experience dimensions.

\paragraph{Toward Task-Level Evaluation.}
Our experiments primarily assess gesture-level prediction accuracy, which provides a meaningful foundation for understanding model capabilities on \hiddeninteractions. However, two aspects of our evaluation warrant extension. First, our test set exclusively contains screens where \hiddeninteractions exist, whereas realistic deployment requires agents to handle mixed conditions—including screens with visible affordances for the same tasks and screens where tasks are infeasible. Second, our accuracy metric evaluates exact gesture matching, yet different gestures can sometimes achieve equivalent functionality; for instance, both double tap and pinch can accomplish zooming, but predicting one when the other is the ground truth is currently marked as incorrect. Future work should address these limitations by developing mixed test sets and evaluation frameworks that account for functional equivalence, enabling more comprehensive assessment of how \ghostui-trained models perform in real-world agent pipelines.
\section{Conclusion}
In this paper, we presented \ghostui, the first dataset explicitly designed to document \hiddeninteractions in mobile user interfaces. Through automated probing of 81 Android applications, we identified and validated 1,970 instances where essential functionality exists without visual affordances—revealing a critical blind spot in current UI understanding approaches. Our experiments demonstrate that training on \ghostui enables vision language models to predict both interaction gestures and resulting UI transitions. These findings highlight the importance of expanding both the action spaces and contextual understanding capabilities of mobile automation systems. By open-sourcing our collection framework alongside the dataset, we provide both the empirical foundation and methodological tools for the research community to address this overlooked dimension of mobile interaction. We hope \ghostui catalyzes further research at the intersection of UI understanding, mobile task automation, and human-computer interaction, ultimately paving the way for more intuitive and capable mobile experiences for all users.

\begin{acks}
This work was supported by the National Research Foundation of Korea (NRF) grant funded by the Korean government (MSIT) (No. 2023R1A2C200520911), the Institute of Information \& Communications Technology Planning \& Evaluation (IITP) grant funded by the Korean government (MSIT) [NO.RS-2021-II211343, Artificial Intelligence Graduate School Program (Seoul National University)], and by the SNU-Global Excellence Research Center establishment project. The ICT at Seoul National University provided research facilities for this study.
\end{acks}

\bibliographystyle{ACM-Reference-Format}

\clearpage
\newcolumntype{M}{>{\centering\arraybackslash}p{1.2cm}}
\onecolumn 

\appendix
\newpage

\section{UI Task Generation Prompt}
\label{appendix:task_generation}
\begin{figure*}[h]
    \centering
    \includegraphics[width=\textwidth]{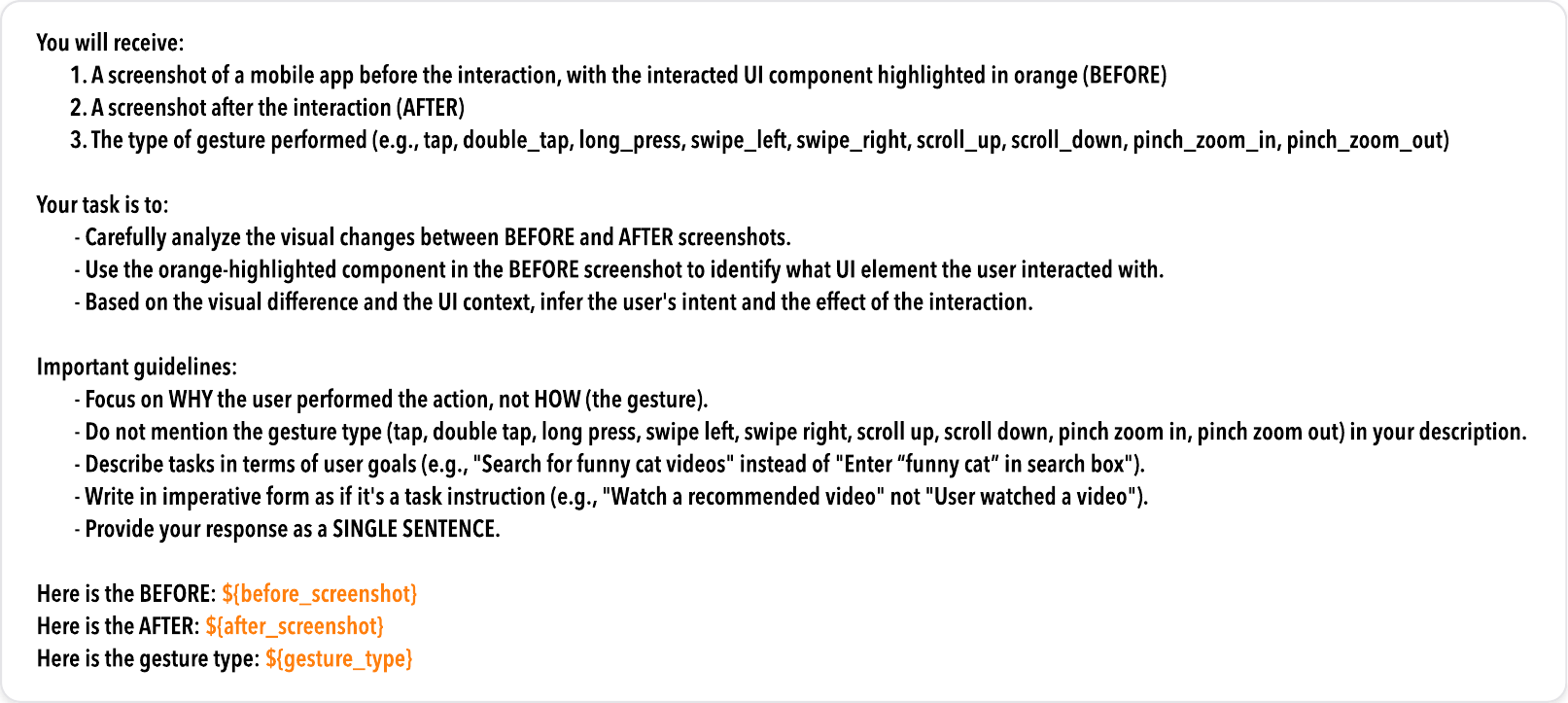}
    \caption{Prompt Template for Task Generation. \textmd{GPT-4o is given before and after screenshots of a mobile app (with the interacted UI element highlighted in the before image) and the gesture type. The task is to generate a single sentence, imperative instruction that captures the user’s intent and the effect of the interaction, without mentioning the gesture.}}
    \label{fig:prompt_task_generation}
\end{figure*}
\clearpage

\section{Experimental Prompts}
\label{appendix:prompts}
\subsection{\HiddenInteraction Prediction}
\label{appendix:hiddeninteraction}
\begin{figure*}[h]
    \centering
    \includegraphics[width=\textwidth]{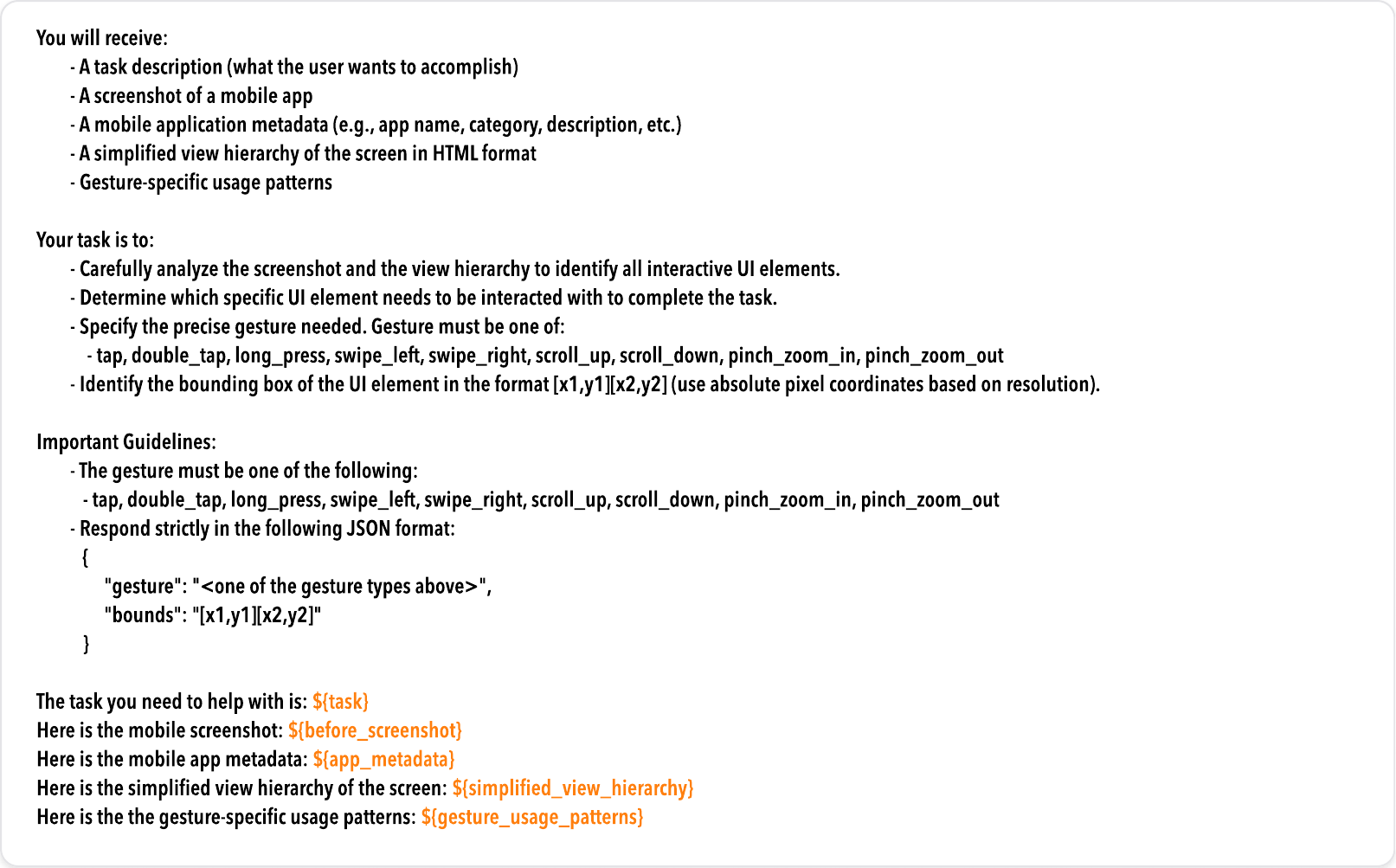}
    \caption{Prompt Template for Hidden Interaction Prediction. \textmd{Models are given a task description, a mobile app screenshot, app metadata, a simplified HTML view hierarchy, and gesture-specific usage patterns. The task is to identify the correct UI element, select an appropriate gesture from a predefined set, and return both the gesture and bounding box in a structured JSON format. An ablation study tests the effect of removing the view hierarchy, gesture patterns, or app metadata to measure each component’s impact on performance.}}
    \label{fig:prompt_hidden_prediction}
\end{figure*}
\clearpage

\subsection{UI Transition Prediction}
\label{appendix:uitransition}
\begin{figure*}[h]
    \centering
    \includegraphics[width=\textwidth]{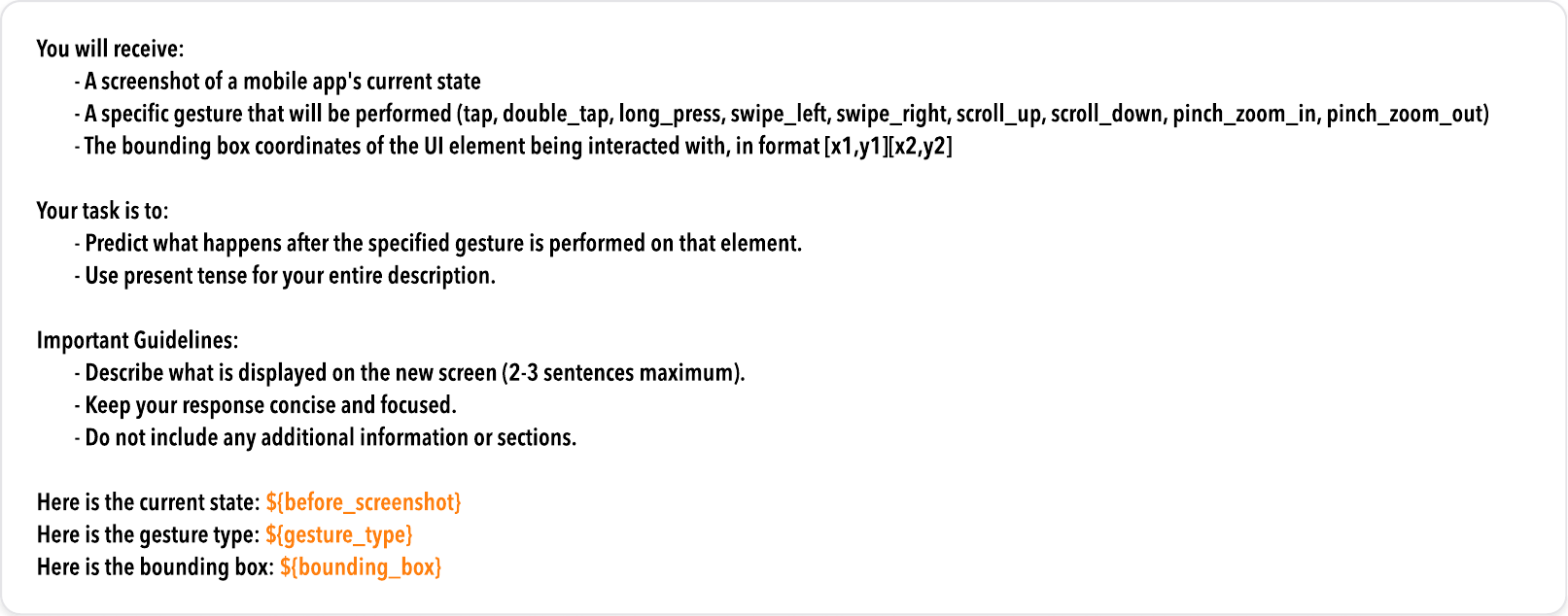}
    \caption{Prompt Template for UI Transition Prediction. \textmd{Models are given a screenshot of a mobile app’s current state, a specific gesture to be performed, and the bounding box of the target UI element. The task is to predict what the screen will display after the gesture is executed on the specified element. The output is a concise 2–3 sentence description in present tense, focusing solely on the visual result.}}
    \label{fig:prompt_groundtruth}
\end{figure*}

\begin{figure*}[h]
    \centering
    \includegraphics[width=\textwidth]{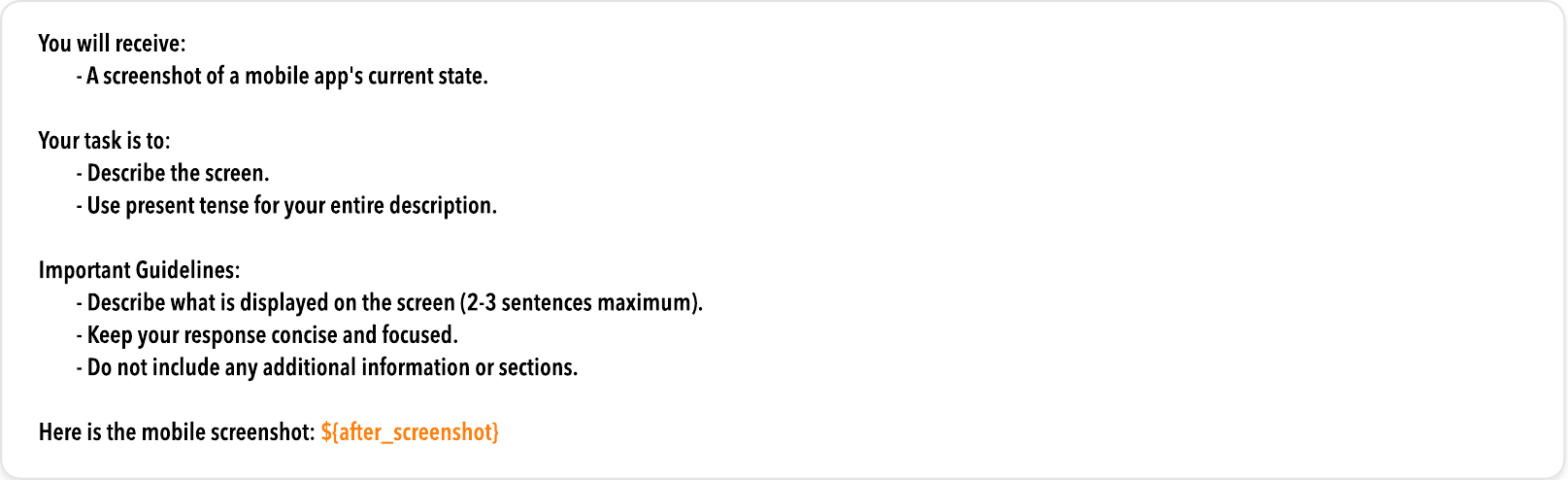}
    \caption{Prompt Template for AFTER Screenshot Description. \textmd{GPT-4o is given a screenshot of a mobile app’s current state. The task is to concisely describe the visible content on the screen using 2–3 sentences in present tense, focusing only on what is shown without adding extra context.}}
    \label{fig:prompt_ui_transition}
\end{figure*}

\end{document}